\documentclass{article}

\usepackage{microtype}
\usepackage{amsmath,amssymb,amsthm}
\usepackage{graphicx}
\usepackage{url}
\usepackage{tikz}
\usepackage{algorithm}
\usepackage{algorithmic}
\usepackage{rotating}
\usepackage{ifpdf}
\ifpdf
  \usepackage{hyperref}
\fi

\newcommand{\Cov}{\text{Cov}}
\newcommand{\Var}{\text{Var}}
\newcommand{\thetab}{\boldsymbol{\theta}}
\newcommand{\etab}{\boldsymbol{\eta}}

\newtheorem{theorem}{Theorem}[section]

\author{Lucas Seiffert\thanks{lucas.seiffert@utdallas.edu}, Felipe Pereira\thanks{luisfelipe.pereira@utdallas.edu}}
\title{A Framework to Analyze Multiscale Sampling MCMC Methods}
\date{\small{\it Department of Mathematical Sciences, The University of Texas at Dallas, Richardson, TX, 75080, USA}}

\begin{document}

\maketitle

\begin{abstract}
We consider the theoretical analysis of Multiscale Sampling Methods, which are a new class of gradient-free Markov chain Monte Carlo (MCMC) methods for high dimensional inverse differential equation problems. A detailed presentation of those methods is given, including a review of each MCMC technique that they employ. Then, we propose a two-part framework to study and compare those methods. The first part identifies the new corresponding state space for the chain of random fields, and the second assesses convergence conditions on the instrumental and target distributions. Three Multiscale Sampling Methods are then analyzed using this new framework. 
\end{abstract}

%%%%%%%%%%%%%%%%%%%%%%%%%%%%%%%%%%%%%%%%%%%%%%%%%%%%%%%%%%%%%%%%%%%%%%%%%%%%%%%%%%%%
%%%%%%%%%%%%%%%%%%%%%%%%%%%%%%%%%%%%%%%%%%%%%%%%%%%%%%%%%%%%%%%%%%%%%%%%%%%%%%%%%%%%
\section{Introduction} 
\label{sec:intro}

The Bayesian approach to inverse PDE problems has received considerable attention in the recent years, especially when the task is their uncertainty quantification \cite{stuart2010inverse, kaipio2005statistical}.
Sampling from the posterior distributions that arise in those problems usually requires Markov Chain Monte Carlo (MCMC) methods, and the simulation of large, realistic problems, is still a very challenging task: each iteration of the Markov chain must solve an expensive partial differential equation. 

For example, a fundamental problem in the area of subsurface flows is how to obtain the rock properties of a porous medium given some measurements of the flow that passes through it. Because those properties are highly heterogeneous, a very fine grid is required to simulate the problem, which may result in billions of points in a numerical grid \cite{jaramillo2022towards}. 

Multiscale Sampling Methods (MSM) are a new class of derivative-free MCMC methods that have shown promising experimental results, with a significant improvement in convergence \cite{ali2024multiscale}. They take inspiration from domain decomposition methods in numerical analysis: sampling is performed locally in disjoint subdomains, and different techniques are possible to couple those local fields into a global one. Notice that our use of the term \textit{coupling} has no connection with the same term in the theoretical study of Markov chains.

The \textit{local sampling} of our method is made possible by a combination of different MCMC techniques: each subdomain has its own basis of functions, generated by a dimension reduction technique such as the Karhunen-Lo\`eve decomposition. Then, a Metropolis-within-Gibbs MCMC approach keeps all subdomains frozen but one, during each iteration. That is, at each iteration, the sampled field changes in only one of the subdomains. Next, a \textit{coupling technique} removes discontinuities from this new sampled field, before the likelihood is calculated. Finally, an acceptance/rejection step like the Metropolis-Hastings or Two-Stage Delayed Acceptance method is performed. 

This paper is the first step into the theoretical study of Multiscale Sampling Methods. Typically, a new MCMC method is analyzed from the point of view of convergence of its Markov chain, and then an error analysis is performed to determine the expected error between the simulated and the true distribution \cite{dodwell2019multilevel}. 

Our first goal is therefore to check whether the Multiscale Sampling methods are well-defined, in the sense that their Markov chain has the desired limiting distribution. For this, we first describe each technique mentioned above, along with their criteria for convergence. Then, we show where each component fits into our MCMC. 

However, when it comes to the new Multiscale Sampler Method, we are faced with a new challenge. Namely, we shall see that the coupling step, which generates the global permeabilty field from the local samples, affects the method in ways that are unusual to typical MCMC. The crucial questions here are: ``if I couple the local fields in a different way, what happens? Where will this modification go?''. 

We propose a new framework to answer those questions. That is, we show how different coupling strategies modify different parts of the MCMC, the most important being which function space we are effectively sampling from. In this way, this analysis will also pave way to inspire multiscale sampling methods. Finally, some issues we encountered are left as future work. 

Our contributions in this paper can be summarized as:

\begin{itemize}
\item The Multiscale Sampling Method is explained in detail, including all the Markov Chain Monte Carlo techniques it is based on.
\item A new framework to study Multiscale Sampling Methods is introduced, with two major goals. First, to confirm that the method is a correct MCMC. And second, to compare different Multiscale Sampling Methods.
\item This framework is applied to three different Multicale Sampling Methods, including the original method in \cite{ali2024multiscale}. 
\end{itemize}

\vspace{0.5cm}
The paper is organized as follows. In Section \ref{sec:background}, we review the existing techniques from which the Multiscale Sampling Method is derived, and enumerate the conditions that must be satisfied to ensure their convergence. Section \ref{sec:inverseproblem} introduces an inverse elliptic problem that motivated the development of the Multiscale Sampling Method, to be described in Section \ref{sec:multiscale}. Then, we explain in Section \ref{sec:framework} our new framework to analyze this family of methods, and apply it to different Multiscale Sampling Methods in Section \ref{sec:analysis}, highlighting the challenges that come with each coupling decision.

%%%%%%%%%%%%%%%%%%%%%%%%%%%%%%%%%%%%%%%%%%%%%%%%%%%%%%%%%%%%%%%%%%%%%%%%%%%%%%%%%
%%%%%%%%%%%%%%%%%%%%%%%%%%%%%%%%%%%%%%%%%%%%%%%%%%%%%%%%%%%%%%%%%%%%%%%%%%%%%%%%%
\section{Some techniques in Markov chain Monte Carlo} 
\label{sec:background}

The goal of this section is to describe the MCMC methods and ideas that are used by the Multiscale Sampling Method, and there are no new results. Some key calculations are done in more detail than in the references cited, to confirm what assumptions they require.

Some conditions for convergence of a Markov chain will be enumerated and commented on, but precise definitions would require a thorough discussion on Markov chain theory. Instead, we refer to the authoritative references \cite{levin2017markov} for chains with finite state spaces, and \cite{meyn2009markov} for chains with general state spaces. 

The last topic is an introduction to the Karhunen-Lo\`eve decomposition of a stochastic field, which is the technique we chose for dimensional reduction of the problem. This is the starting point for much of the discussion about the function spaces in Multiscale Sampling Methods, and it shows why the analysis of a typical MCMC method is more straightforward than in multiscale sampling. 

%%%%%%%%%%%%%%%%%%%%%%%%%%%%%%%%%%%%%%%%%%%%%%%%%%%%%%%%%%%%%%%%%%%%%%%%%%%%%%%%%%%%%%%%%
\subsection{The Metropolis-Hastings method}
\label{sec:mh}

This is the pioneering method that enabled the simulation of distributions that were previously intractable, and it is the basis to many of the modern MCMC methods. This subsection is based entirely on the textbook of \cite{casella2002statistical}, where the state space of the Markov Chain is assumed to be finite-dimensional. In theory, we shoud be sampling from an infinite dimensional function space. But in practice it is a finite dimensional state space, due to the discretization of the domain of the differential equation. The study of conditions on the instrumental distributions for more general state spaces in a Metropolis-Hastings method is provided in \cite{tierney1998note}.  

Denote by $f$ the probability density to be simulated. It is sometimes called the \textit{target distribution}. New sample values are generated according to an \textit{instrumental} (or \textit{proposal}) distribution $q(y|x)$. The success of the Metropolis-Hastings method can be attributed to the considerable freedom to choose such $q$, and in the simplicity of the algorithm, which is summarized in Algorithm \ref{alg:mh}. Since we're doing Bayesian analysis, it is convenient to use $f(x) \propto L(x)\pi(x)$, where $L$ is the likelihood function applied to the parameter $x$ and $\pi$ is its prior density. 

\begin{algorithm}
\caption{Metropolis-Hastings MCMC}
\label{alg:mh}
\begin{algorithmic}[1]
\STATE{Given $x^{(t)}$}
\STATE{Draw $y^{(t)}\sim q(\cdot|x^{(t)})$}
\STATE{Set $x^{(t+1)}=y^{(t)}$ with probability $\alpha(x^{(t)}, y^{(t)})$, 
\[\alpha(x,y)=\min\left\{1,\frac{f(y)}{f(x)}\frac{q(x|y)}{q(y|x)}\right\}=\min\left\{1,\frac{L(y)\pi(y)}{L(x)\pi(x)}\frac{q(x|y)}{q(y|x)}\right\},\] and set $x^{(t+1)}=x^{(t)}$ otherwise}
\STATE{Set $t\leftarrow t+1$ and go to Step 1}
\end{algorithmic}
\end{algorithm}

The Markov chain $\{x^{(t)}\}_t$ generated by the method has transition kernel
\[K(x, y) = \rho(x,y)q(y|x)+(1-r(x))\delta_x(y),\]
with $r(x) = \int\rho(x,y)q(y|x)dy$. This kernel is used in proofs for the convergence of the Metropolis-Hastings approach, and later for the convergence of Two-Stage Delayed Acceptance methods.

There is one condition on $f$ and two conditions on $q$ that must be checked to ensure that $\{x^{(t)}\}_t$ is a Markov Chain that converges to $f$, as desired. The notation $\text{supp}\ f$ is used for the support of $f$. 

\vspace{0.5cm}
\noindent\textbf{Condition 1.} The set $\text{supp}\ f$ must be connected. Otherwise, the method must be modified to sample from each connected component. 

\vspace{0.2cm}
\noindent\textbf{Condition 2.} The support of $q$ contains the support of $f$, 
\[\bigcup_{x\in\text{supp}\ f}\text{supp}\ q(\cdot|x)\supset \text{supp}\ f.\]

\vspace{0.2cm}
\noindent\textbf{Condition 3.} \textit{Positivity} property on $q$:
\[q(y|x) > 0 \text{ for every }(x,y)\in\text{supp}\ f\times\text{supp}\ f.\]

\vspace{0.2cm}
\noindent\textbf{Condition 4.} There is a nonzero probability that a step will be rejected.

\vspace{0.2cm}
\noindent\textbf{Alternative to condition 3.} The function $f$ must be bounded and positive on every compact set of its support, and there exist positive numbers $\epsilon$ and $\delta$ such that \[q(y|x)>\epsilon\quad \text{if }|x-y|<\delta.\] 

\vspace{0.5cm}
\textit{Condition 3} is a sufficient condition for the chain to be $f$-irreducible. Informally, being $f$-irreducible means that every measurable subset $A$ of the state space such that $\int_Af(x)dx >0$ will have nonzero probability of being reached by the chain in finite time, and this should happen for any starting point for the chain.   

\textit{Condition 4} guarantees that the chain is aperiodic, that is, that the state space cannot be partitioned into cycles that prevent the chain from returning to the same point in one step. 

The last condition can substitute 3 and 4 to show $f$-irreducibility and aperiodicity and may be useful for the Multiscale Sampling Methods.  

The following theorems put everything together. They prove that, by using the probability of acceptance $\alpha(\cdot, \cdot)$ in Algorithm \ref{alg:mh}, the detailed balance condition is satisfied for $f$ and that $f$ is the stationary distribution. %\cite{casella2002statistical}. % page 272

\begin{theorem}[\cite{casella2002statistical}, Theorem 7.2]\label{thm:detailedbalance}
Let $(X^{(t)})$ be the chain produced by the Metropolis-Hastings algorithm. For every conditional distribution $q$ whose support includes the support of $f$, 

(a) the kernel $K$ of the chain satisfies the detailed balance condition with $f$, that is,
\[K(y,x)f(y) = K(x,y)f(x)\quad\text{for all }(x,y).\]

(b) the density $f$ is a stationary distribution of the chain.
\end{theorem}

\begin{theorem}[\cite{casella2002statistical}, Theorem 7.4]\label{thm:ergodic}
Suppose that $\{x^{(t)}\}$ generated by the Metropolis-Hastings method is $f$-irreducible and aperiodic. Then it is convergent, in the following sense.

(i) If $h$ satisfies $\int h(x)f(x)dx < \infty$, then
\[\lim_{T\rightarrow\infty}\frac{1}{T}\sum_{t=1}^Th(x^{(t)})\rightarrow\int h(x)f(x)dx\quad a.e.\text{ in } f.\]
That is, the sum of random variables on the left-hand side converges everywhere in the state space except for perhaps a measurable subset $A$ such that $\int_A f(x)dx=0$.

(ii) For every initial distribution $\mu$, and denoting by $K^n(x,\cdot)$ the kernel of the chain for $n$ transitions, then
 \[\lim_{n\rightarrow\infty}\bigg|\bigg|\int K^n(x,\cdot)\mu(dx)-f\bigg|\bigg|_{TV} = 0,\]
where $||\cdot||_{TV}$ is the total variation norm given by
 \[||\mu_1-\mu_2||_{TV}=\sup_A|\mu_1(A)-\mu_2(A)|.\] 
\end{theorem}

%%%%%%%%%%%%%%%%%%%%%%%%%%%%%%%%%%%%%%%%%%%%%%%%%%%%%%%%%%%%%%%%%%%%%%%%%%%%%%%%%%%%%%%%%%%%%
\subsection{The Two-Stage Delayed Acceptance MCMC}
\label{sec:twostage}

This a modification of the Metropolis-Hastings algorithm for cases where the target probability density $f$ is expensive to be computed, and it was introduced in \cite{christen2005markov}. The method is particularly useful in Bayesian methods whose likelihoods that require the numerical approximation to the solution of a differential equation. As hinted by the name given to this method, an extra step is added to the Metropolis-Hastings algorithm, in order to ``delay'' the expensive evaluation of $f$ on a new sample value to only those values that have more chances of being accepted. 

Those ideas are summarized in Algorithm \ref{alg:twostage}. Our description will follow the original paper of \cite{christen2005markov}, with some adaptations to keep the notation in this work consistent. When a new sample value is drawn from the instrumental distribution $q$, it will go through the step in line 4 in the algorithm. There, a coarse approximation $f_c$ to $f$ will be used to determine whether this sample value can be promoted to the Metropolis-Hastings step in line $6$. Otherwise, the new value is rejected, and the chain keeps its old value for iteration $t+1$. If the value is promoted, the \textit{fine} calculation of $f$, denote by $f_f$, is performed. It is used to finally evaluate if this new value will be accepted by the chain as in line 6 of the algorithm.

\begin{algorithm}
\caption{Two-Stage Delayed Acceptance MCMC}
\label{alg:twostage}
\begin{algorithmic}[1]
\STATE{Given $x^{(t)}$}
\STATE{Draw $y^{(t)}\sim q(\cdot|x^{(t)})$}
\STATE{Evaluate the coarse approximation $f_c(y^{(t)})$}
\STATE{Promote $y^{(t)}$ with probability $g(x^{(t)}, y^{(t)})$,
\[g(x,y)=\min\left\{1,\frac{f_c(y)}{f_c(x)}\frac{q(x|y)}{q(y|x)}\right\},\] otherwise set $x^{(t+1)}=x^{(t)}$ and start a new iteration at Step 1}
\STATE{Evaluate the fine approximation $f_f(y^{(t)})$}
\STATE{Set $x^{(t+1)}=y^{(t)}$ with probability $\rho(x^{(t)}, y^{(t)})$,
\[\rho(x,y) =\min\left\{1,\frac{q^*(x|y)}{q^*(y|x)}\frac{f_f(y)}{f_f(x)}\right\},\]
where $q^*(y|x) = g(x,y)q(y|x)$ when $x\neq y$. Set $x^{(t+1)}=x^{(t)}$ otherwise}
\STATE{Set $t\leftarrow t+1$ and go to Step 1}
\end{algorithmic}
\end{algorithm}

In a Bayesian context, we can denote $f_c(x) \propto L_c(x)\pi(x)$ and $f_f(c) \propto L_f(x)\pi(x)$, indicating that the approximations will be done on the likelihood function. And in practice, the expressions for the acceptance probabilities $g$ and $\rho$ can be simplified. The case for $g$ will be explained in Section \ref{sec:pCN}, and a theorem for $\rho$ is shown below. This result is not new and it's how the method is typically coded \cite{ali2021multiscale}. The proof that follows is simple and written by us from scratch.

\begin{theorem}
For any chosen instrumental distribution $q$, the probability of acceptance $\rho$ can be simplified and written as
\[\rho(x,y) = \min\left\{1, \frac{f_f(y)}{f_f(x)}\frac{f_c(x)}{f_c(y)}\right\}=\min\left\{1, \frac{L_f(y)}{L_f(x)}\frac{L_c(x)}{L_c(y)}\right\}\]
\end{theorem}

\begin{proof}
Our goal is to simplify the expression 
\[\frac{q^*(x|y)}{q^*(y|x)}\frac{f_f(y)}{f_f(x)},\]
and the proof is divided into two cases.

(i) Suppose $\frac{f_c(y)q(x|y)}{f_c(x)q(y|x)} < 1$. This will imply 
\[g(y,x) = 1,\qquad\text{and}\qquad g(x,y) = \frac{f_c(y)q(x|y)}{f_c(x)q(y|x)}.\]
Also,
\[q^*(y|x) = g(x,y)q(y|x), \qquad\text{and}\qquad q^*(x|y) = q(x|y),\]
and $\rho$ becomes
\begin{align*}
\rho(x,y) &= \min\left\{1, \frac{q(x|y)f_f(y)}{g(x,y)q(y|x)f_f(x)}\right\}
=\min\left\{1, \frac{q(x|y)f_f(y)}{\frac{f_c(y)q(x|y)}{f_c(x)q(y|x)}q(y|x)f_f(x)}\right\}\\
&=\min\left\{1, \frac{f_c(x)f_f(y)}{f_c(y)f_f(x)}\right\}
\end{align*}

Substituting $f_c(x) = L_c(x)\pi(x)$ and $f_f(c) = L_f(x)\pi(x)$ yields the second assertion of our theorem. 

(ii) The second case is when $\frac{f_c(y)q(x|y)}{f_c(x)q(y|x)}\geq 1$. Analogously to the first case,
\[g(x,y) = 1,\qquad\text{and}\qquad g(y,x) =  \frac{f_c(x)q(y|x)}{f_c(y)q(x|y)},\]
and
\[q^*(y|x) = q(y|x),\qquad\text{and}\qquad q^*(x|y) = g(y,x)q(x|y).\]
Then,
\begin{align*}
\rho(x,y) &= \min\left\{1, \frac{g(y,x)q(x|y)f_f(y)}{q(y|x)f_f(x)}\right\}
=\min\left\{1, \frac{\frac{f_c(x)q(y|x)}{f_c(y)q(x|y)}q(x|y)f_f(y)}{q(y|x)f_f(x)}\right\}\\
&=\min\left\{1, \frac{f_c(x)f_f(y)}{f_c(y)f_f(x)}\right\}
\end{align*}
The second assertion will also follow from a simple substitution. 
\end{proof}

The proof that this method converges to $f$ is given in Theorem \ref{thm:twostage}. It is based on the theorems for the Metropolis-Hastings method, so some conditions will coincide with the ones listed in the previous subsection section. 

Those new conditions are

\vspace{0.5cm}
\noindent\textbf{Condition 5.} $q(y|x)>0$ implies $f_c(y)>0$,

\vspace{0.5cm}
\noindent\textbf{Condition 6.} the transition kernel $K_q$ given by a Metropolis-Hastings with $q$ as an instrumental distribution is reversible. In this context, reversibility is equivalent to the detailed balance condition from Theorem \ref{thm:detailedbalance} from the previous subsection.

\vspace{0.5cm}
The convergence theorem for this method is as follows.

\begin{theorem}[\cite{christen2005markov}]\label{thm:twostage}
If the Metropolis-Hastings algorithm with $q$ as a proposal (kernel $K_q(\cdot,\cdot)$) is f-irreducible, $q$ is reversible and $q(y|x)>0$ implies $f_c(y)>0$, then 

(a) $f$ is an invariant distribution for $K$ and $K$ is f-irreducible

(b) Moreover, $K_q(x,x)>0\implies K(x,x)>0$ for any $x$, and the resulting chain is strongly aperiodic. 
\end{theorem}

The ergodic theorem \ref{thm:ergodic} can then be applied to give the convergence of the method. 

Condition 6 is used in the proof of Theorem \ref{thm:twostage} to show that $f$ is an invariant distribution to this new chain. As an alternative, let us try to apply Theorem \ref{thm:detailedbalance} from the previous subsection to reach the same conclusion. The requirements are Condition 2, Condition 5 and the new 

\vspace{0.5cm}
\noindent\textbf{Condition 6'.} For any $x$ and $y$ in the state space of the Markov chain, 
\[q(y|x)>0\implies q(x|y)>0.\]
Note that this is trivially satisfied if Condition 3 is true for $q$. 

Then, 
\begin{theorem}\label{thm:tscondition6}
Under conditions 2, 5 and 6' for $q$, the conditional distribution $q^*$ of Algorithm \ref{alg:twostage} contains the support of $q$.
\end{theorem}
\begin{proof}
The case when $x=y$ is trivial, so we consider $y\neq x$. Moreover, if $x$ is part of the chain, then we must have that $f_c(x)>0$ by Condition 5. Then
\[q^*(y|x) = g(x,y)q(y|x) = q(y|x)\min\left\{1, \frac{q(x|y)f_c(y)}{q(y|x)f_c(x)}\right\}.\]
By the conditions stated, if $q(y|x)>0$, then all the terms in the equality above are non negative. Therefore, $q^*(y|x)>0$ as required.
\end{proof}

Condition 3 is now satisfied for $q^{*}$ if it is already satisfied for $q$, and Theorem \ref{thm:detailedbalance} is applied to the instrumental distribution from the Two-Stage Delayed Acceptance method.

%%%%%%%%%%%%%%%%%%%%%%%%%%%%%%%%%%%%%%%%%%%%%%%%%%%%%%%%%%%%%%%%%%%%%%%%%%%%%%%%%%%%%%%%%%
\subsection{The preconditioned Crank-Nicolson step}
\label{sec:pCN}

In the previous two subsections, we presented general MCMC methods that leave open the choice of the \textit{instrumental distribution} $q$. A simple idea is to perform a random walk on the state space, and this has the name of Random Walk Sampler (RWS) \cite{casella2002statistical}. A new value $y$ is generated by simulating $\xi\sim N(0, \mathcal{C})$ and setting
\[y = x +\beta\xi,\]
where $\beta$ is parameter to be chosen. This means that 
\[q_{RWS}(\cdot,x)\sim N(x, \beta^2\mathcal{C}),\]
with some covariance matrix $\mathcal{C}$. 

Besides being simple to compute, this choice of $q$ has another advantage: the probability of acceptance in the Metropolis-Hastings method simplifies to 
\[\alpha(x,y) = \min\left\{1, \frac{f(y)}{f(x)}\right\},\]
since
\[\frac{q_{RWS}(x|y)}{q_{RWS}(y|x)}=\exp\left((x-y)^T\mathcal{C}^{-1}(x-y)-(y-x)^T\mathcal{C}^{-1}(y-x)\right)=1.\]

The preconditioned Crank-Nicolson (pCN) step is an alternative to the Random Walk Sampler, and it is motivated by Langevin dynamics \cite{cotter2011variational, cotter2013mcmc}. It comes from the discretization of a particular stochastic differential equation that has the posterior distribution $f$ of exponential form $\exp(\Phi(x))$ for some potential function $\Phi$ or a Gaussian prior as an invariant measure. This discretization generates a Markov Chain that itself may already converge to the posterior distribution, but it is fed into a Metropolis-Hastings method to fix any deviations due to discretization errors. 

There are two things to be decided: a parameter $\beta$ that decides the size of the step to be taken, and the covariance matrix $\mathcal{C}$ for the distribution of the step increment. With this, the preconditioned Crank-Nicolson step reads

Draw $\xi\sim N(0,\mathcal{C})$, and set 
\[y = (1-\beta^2)^{\frac{1}{2}}x+\beta\xi.\]

Since this is $y$ is an affine transformation of a Gaussian random variable, its conditional distribution is given by
\[q_{pCN}(y|x)\sim N((1-\beta^2)^{\frac{1}{2}}, \beta^2\mathcal{C}).\] 

As it is the case with the Random Walk Sampler, the use of a preconditioned Crank-Nicolson step may simplify the probability of acceptance $\alpha$ of the Metropolis-Hastings method. However, we emphasize that this only happens if the prior has a particular structure, and this will be summarized in a theorem below. The result is used in the papers that mention the pCN, but without proof. 

\begin{theorem}\label{thm:pCN}
For any $\beta$ chosen, 
\[\frac{q_{pCN}(x|y)}{q_{pCN}(y|x)} = \frac{\pi(x)}{\pi(y)},\]
where $\pi\sim N(0,\mathcal{C})$.
\end{theorem}

\begin{proof}

First, we notice that
\[\frac{q_{pCN}(x|y)}{q_{pCN}(y|x)} = \exp\left(-\frac{1}{2\beta^2}A\right),\]
where
\begin{align*}A=(&x-\sqrt{1-\beta^2}y)^T\mathcal{C}^{-1}(x-\sqrt{1-\beta^2}y)-(y-\sqrt{1-\beta^2}x)^T\mathcal{C}^{-1}(y-\sqrt{1-\beta^2}x)\\
&= x^T\mathcal{C}^{-1}x+(1-\beta^2)y^T\mathcal{C}^{-1}y-2\sqrt{1-\beta^2}x^T\mathcal{C}^{-1}y-y^T\mathcal{C}^{-1}y-(1-\beta^2)x^T\mathcal{C}^{-1}x+\\
&\qquad+2\sqrt{1-\beta^2}x^T\mathcal{C}^{-1}y\\
&=\beta^2(x^T\mathcal{C}^{-1}x-y^T\mathcal{C}^{-1}y)
\end{align*}

Therefore, 
\[\frac{q_{pCN}(x|y)}{q_{pCN}(y|x)} = \exp\bigg(-\frac{1}{2\beta^2}\beta^2(x^T\mathcal{C}^{-1}x-y^T\mathcal{C}^{-1}y)\bigg) = \frac{\pi(x)}{\pi(y)},\]
\end{proof}

Now, suppose that $f(x)\propto L(x)\pi(x)$, with $\pi\sim N(0,\mathcal{C})$ as in the theorem. Then the probability of acceptance $\alpha$ in the Metropolis-Hastings method simplifies to
\[\alpha(x,y) = \min\left\{1,\frac{L(y)\pi(y)q(x|y)}{L(x)\pi(x)q(y|x)}\right\}\min\left\{1,\frac{L(y)}{L(x)}\right\}.\]

When it comes to convergence of the MCMC using those instrumental distributions, it is not difficult to show that if the density $f$ is strictly positive in the whole state space, then \textit{Conditions 1-6} are satisfied \cite{tierney1994markov}. This will be explored when we perform the analysis of the Multiscale Sampling Methods later.

%%%%%%%%%%%%%%%%%%%%%%%%%%%%%%%%%%%%%%%%%%%%%%%%%%%%%%%%%%%%%%%%%%%%%%%%%%%%%%%%%%%%
\subsection{The Metropolis-within-Gibbs}
\label{sec:MwG}
% 2D Gibbs: RC p. 371

Let $x=(x_1, x_2,\ldots, x_D)$ be the vector of parameters for the distribution $f$ to be simulated. Let us also divide the indices $1, 2,\ldots, D$ into blocks $B_1, \ldots, B_l$. If we can find a way to simulate from the conditional distributions
\[f(x_{B_i}|x_{B_1},x_{B_2},\ldots, x_{B_{i-1}}, x_{B_{i+1}},\ldots, x_{B_l}),\]
then the \textit{Gibbs Sampler} method can be used as an alternative to the Metropolis-Hastings MCMC \cite{casella2002statistical}. A step of the Gibbs Sampler can be written as a \textit{cycle} that simulates from one block of parameters $B_i$ at a time, and it is summarized in Algorithm \ref{alg:gibbs}.

\begin{algorithm}
\caption{Gibbs Sampler}
\label{alg:gibbs}
\begin{algorithmic}[1]
\STATE{Given $x^{(t)}$}
\STATE{Draw $x^{(t+1)}_{B_1}\sim f(\cdot_{B_1}|x_{B_2}^{(t)}, x_{B_3}^{(t)},\ldots, x_{B_l}^{(t)})$}
\STATE{Draw $x^{(t+1)}_{B_2}\sim f(\cdot_{B_2}|x_{B_1}^{(t)}, x_{B_3}^{(t)}, \ldots, x_{B_l}^{(t)})$}
\STATE{$\ldots$}
\STATE{Draw $x^{(t+1)}_{B_l}\sim f(\cdot_{B_l}|x_{B_1}^{(t)}, x_{B_2}^{(t)}, \ldots, x_{B_{l-1}}^{(t)})$}
\STATE{Go to Step 1 with $t\leftarrow t+1$}
\end{algorithmic}
\end{algorithm}

The Gibbs Sampler is typically chosen when those conditional distributions are easy to compute, but there are also situations when it may be convenient to sample from each block at a time even if those conditionals are not available in closed form. For this, a Metropolis-within-Gibbs approach can be used \cite{muller1991generic,roberts2006harris, sherlock2010random}. It consists of approximating the conditional distribution corresponding to a block $B_i$ by a Metropolis-Hastings step. This step has an instrumental distribution $q_i$ that only changes $x_{B_i}$ and freezes all the other coordinates. 

Conditions to guarantee the convergence of the Gibbs Sampler are discussed in \cite{casella2002statistical} and \cite{tierney1994markov}, and they mention that much of that theory does not directly apply to the Metropolis-within-Gibbs variation. The remaining of this subsection is based on the paper of \cite{roberts2006harris}, which specifically addresses the convergence of Metropolis-within-Gibbs methods.  

Assuming that its Metropolis-Hastings steps satisfy their own convergence conditions, the Metropolis-within-Gibbs will have $f$ as stationary distribution as well. Moreover, the irreducibility property of the chain is satisfied if every direction is sampled from at each cycle. This happens to be our case. 

The difficulty lies in showing that the chain satisfies another property called \textit{Harris recurrence}, which guarantees that the Markov chain converges for any starting point. A few conditions are studied and shown to be sufficient for it to hold \cite{roberts2006harris}. For example, one could show that starting at any point, our method can change each coordinate at least once. 

A condition that can be verified analytically is based on Corollary 18 from \cite{roberts2006harris}:

\vspace{0.5cm}
\noindent\textbf{Condition 7.} The distribution $f$ is such that its $r$-dimensional integral has finite Lebesgue integral over every $r$-dimensional coordinate hyperplane of the state space, for all $1\leq r\leq D$. Here, $D$ is the dimension of the vector of parameters. 

\vspace{0.5cm}
Finally, none of the conditions here can avoid what is the major complication of a Gibbs approach. Namely, many simple examples can be designed showing that the Gibbs method can be \textit{trapped} in practice, and that a good choice of coordinates is crucial for the performance of the method \cite{casella2002statistical}. 

%%%%%%%%%%%%%%%%%%%%%%%%%%%%%%%%%%%%%%%%%%%%%%%%%%%%%%%%%%%%%%%%%%%%%%%%%%%%%%%%%%%%%%%%%%%%%%
\subsection{The Karhunen-Lo\`eve Expansion}
\label{sec:KLE}

Subsections \ref{sec:mh} to \ref{sec:MwG} explained how to simulate a probability distribution $f$ using Markov Chain Monte Carlo. When the state space for $f$ is very large, the simulation can be sped up by performing a dimensional reduction technique. The Karhunen-Lo\`eve expansion (KLE) is one such approach to decompose a random field into a linear combination of basis functions. This representation as an infinite sum can then be truncated given a desired precision for the approximation. 

During the analysis of the Multiscale Sampling Method in Section \ref{sec:analysis}, we'll see that the property of the Karhunen-Lo\`eve Expansion that matters is that this decomposition is a linear combination of some basis functions. Therefore, we give only a brief explanation of this technique and indicate the references \cite{alexanderian2017brief} and \cite{ghanem2003stochastic} for further details. 

To motivate this discussion, let us consider a Bayesian inverse problem as the one in the following Section \ref{sec:inverseproblem}. The state space is a Hilbert space $L^2(\Omega)$ of functions whose  domain is the unit square $\Omega=[0,1]^2$. The goal is to sample from some posterior distribution $f$ on this state space, 
\[f(x)\propto L(x)\pi_{L^2(\Omega)}(x).\]
For example, the prior distribution $\pi_{L^2(\Omega)}$ can be Gaussian with mean 0 and some covariance operator $\mathcal{C}$. The main idea of the Karhunen-Lo\`eve expansion is that this covariance operator can be seen as a positive self-adjoint operator between Hilbert spaces, so the spectral theorem can be applied. This provides us with an orthonormal basis for $L^2(\Omega)$ via the eigenpairs $\{(\lambda_i, \psi_i)\}_{i=1}^{\infty}$. 

Now, any random field $\etab$ over $L^2(\Omega)$ can be decomposed into a linear combination of some new random variables $\{Y_i\}_i$ and the basis functions $\{\psi_i\}_i$,
\[\etab = \sum_{i=1}^{\infty}Y_i\psi_i, \qquad Y_i = \int_{\Omega}\etab(x)\psi(x)dx.\]
The first equality above is actually a theorem of mean-squared convergence of an infinite sum of random variables \cite{ghanem2003stochastic}. 

Two properties of the Karhunen-Lo\`eve expansion are of special interest. First, the truncation up to $N$ terms of this infinite sum will give the best mean-squared approximation of the random field over the finite subspace $V=\text{span}\{\psi_i\}_{i=1}^N$ of $L^2(\Omega)$. And second, if the field is Gaussian, then it can be simulated using independent one-dimensional Gaussian variables $\theta_i$ instead of the more complex $Y_i$ in the previous equation,
\[\etab = \sum_{i=1}^N\sqrt{\lambda_i}\theta_i\psi_i, \qquad \forall i; \theta_i\in N(0,1).\]

It will be convenient in the study of the Multiscale Sampling Methods to introduce one more piece of notation. Assume again that it was decided to truncate the expansion up to $N$ terms. Denoting by $\boldsymbol{\theta}\in\mathcal{R}^N$ the vector of random variables above, the \textit{global assembling operator} $G$ will be the reconstruction of the field from the random vector of parameters $\thetab$:
\begin{align*}
G\colon \mathcal{R}^N&\rightarrow L^2(\Omega)\\
\boldsymbol{\theta}\ &\mapsto\ \etab,
\end{align*}
that is,
\[\etab(x)=G(\thetab)(x)=\sum_{i=1}^m\sqrt{\lambda_i}\psi_i(x)\theta_i,\qquad\forall x\in\Omega.\]

The MCMC simulation will now be performed on the vector $\boldsymbol{\theta}$ instead of the field $\etab$. That, is what we actually simulate is a posterior for $\boldsymbol{\theta}$, 
\[p_{\boldsymbol{\theta}}\propto L(G(\cdot))\pi_{\boldsymbol{\theta}}(\cdot).\]
This is summarized in the Algorithm \ref{alg:kle} below, for a generic MCMC technique. If the prior for $\etab$ is Gaussian, then the theory on the Karhunen-Lo\`eve decomposition suggests that the corresponding prior for $\boldsymbol{\theta}$ be a multivariate standard normal distribution, $\pi_{\boldsymbol{\theta}}\sim N(0, Id_N)$.

\begin{algorithm}
\caption{MCMC approach that generates a chain on the KLE parameters}
\label{alg:kle}
\begin{algorithmic}[1]
\STATE{Given $\boldsymbol{\theta}^{(t)}$}
\STATE{Generate $\boldsymbol{\theta}^*\sim q(\cdot|\boldsymbol{\theta}^{(t)})$}
\STATE{Perform an MCMC acceptance/rejection step on $\boldsymbol{\theta}^*$ using $L(G(\cdot))\pi_{\boldsymbol{\theta}}(\cdot)$ as target distribution, and update $\boldsymbol{\theta}^{(t+1)}\leftarrow\boldsymbol{\theta}^*$ or $\boldsymbol{\theta}^{(t+1)}\leftarrow\boldsymbol{\theta}^{(t)}$ accordingly}
\STATE{Go to Step 1 with $t\leftarrow t+1$}
\end{algorithmic}
\end{algorithm}

What is the relationship between the posterior $p_{\boldsymbol{\theta}}$, on the vector $\boldsymbol{\theta}$, and the posterior of interest $p_{\etab}$, on $\etab$? We can try to understand what is happening in a couple of ways shown below. 

\vspace{0.5cm}
\noindent\textbf{Approach A.} The posterior distribution of $\etab$ is to be interpreted as the probability that the coefficients $\thetab$ of its linear combination of functions are $p_{\thetab}(\boldsymbol{\theta})$. This idea is the backbone of our new framework to analyze the Multiscale Sampling Methods.   

First, note that because of the Karhunen-Lo\`eve approximation, our simulation for $\etab$ is restricted to the subspace 
\[V:=Im(G)=\text{span}\{\psi_1, \ldots, \psi_N\}\]
of $L^2(\Omega)$. This means that we can only compare $p_{\thetab}$ and $p_{\etab}$ for $\etab\in V$. 

Since $\{\psi_1, \ldots, \psi_N\}$ is an orthonormal basis for $V$, any $\widetilde{\etab}\in V$ can be uniquely decomposed into a linear combination of those basis functions. We can denote this process by $G^{-1}(\widetilde{\etab})$. Once this is done, the posterior probability of $\widetilde{\etab}$ is the evaluation of this vector of coefficients on the posterior of $\boldsymbol{\theta}$. That is, 
\[\forall \widetilde{\etab}\in Im(G),\qquad G^{-1}(\widetilde{\etab})=(u_1,u_2, \ldots, u_N)=:\mathbf{u}\]
and we calculate the posterior probability for $\widetilde{\etab}$ based on the probability that the posterior of the coefficients is $\mathbf{u}$,
\[p_{\etab}(\widetilde{\etab})\propto L(\text{data}, G(\mathbf{u}))\pi_{\boldsymbol{\theta}}(\mathbf{u})=L(\text{data},\widetilde{\etab})\pi_{\boldsymbol{\theta}}(G^{-1}(\widetilde{\etab}))\]
We can compare this with the original form for this posterior,
\[p_{\etab}(\widetilde{\etab})\propto L(\text{data},\widetilde{\etab})\pi_{\etab}(\widetilde{\etab}).\]
That is, the relationship between the posteriors of $\etab$ and $\boldsymbol{\theta}$ is, up to the normalizing constant, that the likelihood remains fixed and $\etab$ ``enters'' the prior of $\boldsymbol{\theta}$ via the transformation $G$.  

The importance of this procedure is that, in the case of Multiscale Sampling Methods, we don't know at first what happens to the MCMC method and its target distribution when $G$ is a complicated coupling procedure. However, if the method is still a valid Markov chain in $\thetab$ and there is still a one-to-one correspondence between $\etab$ and $\thetab$, then the ideas above can be used to tell us what we are effectively simulating from. 

This is also a way to compare different multiscale methods: in the variable $\etab$, the likelihood expression for all those methods is the same, and the change will be ``moved'' to the prior $\pi_{\etab}$ via $G$ and $\pi_{\thetab}$. Not only that, but we see that the operator $G$ will modify the finite-dimensional state space $V$ that approximates $L^2(\Omega)$ in the MCMC simulation. That is, each method will also have its own space $V$, and we have a technique to say what this space is.  

\vspace{0.5cm}
\noindent\textbf{Approach B.} The random field $\etab$ can also be described by its first and second moments. This idea is standard in the study of spatial data analysis (\cite{ossiander2014conditional, schabenberger2017statistical}, or the more advanced \cite{christakos1992random}). Those will be written in terms of $p_{\thetab}$, and they can provide some insights in the analysis of the Multiscale Sampling Methods.
\[E_{p_{\etab}}[\etab(x)] = E_{p_{\boldsymbol{\theta}}}\left[\sum_{i=1}^N\sqrt{\lambda_i}\psi_i(x)\theta_i\right] = \sum_{i=1}^N\sqrt{\lambda_i}\psi_i(x)E_{p_{\theta_i}}[\theta_i],\]

\begin{align*}
\Cov_{p_{\etab}}(\eta(x_1),\eta(x_2)) &= \Cov_{p_{\boldsymbol{\theta}}}\left(\sum_{i=1}^N\sqrt{\lambda_i}\psi_i(x_1)\theta_i, \sum_{j=1}^N\sqrt{\lambda_j}\psi_j(x_2)\theta_j\right) \\
&= \sum_{i=1}^N\sum_{j=1}^N\sqrt{\lambda_i\lambda_j}\psi_i(x_1)\psi_j(x_2)\Cov_{p_{\boldsymbol{\theta}}}(\theta_i,\theta_j)\\
&= \sum_{i=1}^N\lambda_i\psi_i(x_1)\psi_i(x_2)c_{ii}+2\sum_{i>j}\sqrt{\lambda_i\lambda_j}\psi_i(x_1)\psi_j(x_2)c_{ij},
\end{align*}
where $c_{ij}$ denote the entries of the posterior covariance matrix for $\boldsymbol{\theta}$.

%%%%%%%%%%%%%%%%%%%%%%%%%%%%%%%%%%%%%%%%%%%%%%%%%%%%%%%%%%%%%%%%%%%%%%%%%%%%%%%%%%
%%%%%%%%%%%%%%%%%%%%%%%%%%%%%%%%%%%%%%%%%%%%%%%%%%%%%%%%%%%%%%%%%%%%%%%%%%%%%%%%%%
\section{A Bayesian Elliptic Inverse Problem}
\label{sec:inverseproblem}

As it was mentioned in Section \ref{sec:intro}, the design of the Multiscale Sampling Method was motivated by inverse problems in subsurface flows. Many of those are time-dependent, but a popular stationary problem that is still useful in practice uses the elliptic model below \cite{bear1987modeling}. It corresponds to a horizontal flow, and its governing equations can be written solely in terms of the fluid pressure $p$. This differential equation depends on the scalar parameter field $\kappa\colon\Omega\rightarrow\mathcal{R}$, which is to be interpreted to be the \textit{permeability field} of the porous medium. 

\[\left\{\begin{array}{l l}
 \nabla\cdot(\kappa\nabla p) = 0&\text{in }\Omega \\
 p = 0 &\text{on }\Omega_L\\
 p = 1 &\text{on }\Omega_R\\
 (\kappa\nabla p) \cdot \mathbf{n} = 0 &\text{on }\Omega_{TB}
 \end{array}\right.\]

\begin{figure}[h!]
\centering
\includegraphics[scale=0.8]{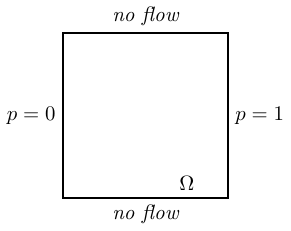}
\caption{Problem domain}
\label{fig:domain}
\end{figure}

For the purposes of this work, the domain $\Omega$ is a one-by-one square as depicted in Figure \ref{fig:domain}. The boundary conditions applied to this model are no-flow Neumann conditions on the top and bottom edges $\Omega_{TB}$, and Dirichlet conditions on the left edge $\Omega_L$ and right edge $\Omega_R$.

Denote the \textit{forward operator} corresponding the above elliptic problem by 
\[F\colon \kappa\mapsto p.\]
It outputs the pressure field $p\in L^2(\Omega)$ of the flow that solves the differential equation given the parameter $\kappa\in L^{\infty}(\Omega)$. Its \textit{inverse} counterpart is to obtain $\kappa$ given measurements $p_{\text{data}}$ of the pressure. Choosing an additive error model for those measurements, we have that 
\[ p_{\text{data}} = F(\kappa)+\boldsymbol{\epsilon},\]
with $\boldsymbol{\epsilon}\in N(0,\Sigma)$.

Since the permeability field $\kappa$ must be nonnegative, it is more convenient to work with its logarithm instead, denoted in this work by $\etab$, 
\[\etab = \log(\kappa).\]

In the Bayesian approach, a prior $\pi(\etab)$ for $\etab$ is chosen, and the measured data $p_{\text{data}}$  is incorporated into a posterior distribution using Bayes' Theorem,
\[\pi(\etab|p_{\text{data}}) \propto L(p_{\text{data}},\etab)\pi(\etab).\]
The likelihood function $L$ is determined by the additive noise model above. That is, 
\[(p_{\text{data}}-F(e^{\etab}))\in N(0,\Sigma),\]
and so
\[L(p_{\text{data}},\etab) \propto \exp\left\{-(p_{\text{data}}-F(e^{\etab}))^T\Sigma^{-1}(p_{\text{data}}-F(e^{\etab}))\right\}.\]

Thanks to the logarithmic tansformation, the prior $\pi(\etab)$ is typically chosen to be a Gaussian field with mean zero and covariance structure
\[R[(x_1, y_1), (x_2, y_2)] = \sigma_{\etab}^2\exp\left(-\frac{|x_1-x_2|^2}{2L^2_x}-\frac{|y_1-y_2|^2}{2L^2_y}\right),\]
where $L_x$ and $L_y$ are the correlation lengths and are what the different scales in the Multiscale Sampling Method will come from. The parameter $\sigma_{\etab}^2$ is the desired variance of the model. 
 
%%%%%%%%%%%%%%%%%%%%%%%%%%%%%%%%%%%%%%%%%%%%%%%%%%%%%%%%%%%%%%%%%%%%%%%%%%%%%%%%%%
%%%%%%%%%%%%%%%%%%%%%%%%%%%%%%%%%%%%%%%%%%%%%%%%%%%%%%%%%%%%%%%%%%%%%%%%%%%%%%%%%% 
\section{The family of Multiscale Sampling Methods}
\label{sec:multiscale}

The Multiscale Sampling Method (MSM) was inspired by multiscale methods for elliptic differential equations in subsurface flows \cite{ali2024multiscale}. Given the high heterogeneity of the rock properties, very fine grids are required for their numerical methods to be sufficiently accurate. To make such simulation more computationally efficient, a multiscale method in numerical analysis will use a domain decomposition approach and partition the domain $\Omega$ into non-overlapping subdomains $\Omega_i$. Inside those subdomains, a local differential equation problem is solved over a finely discretized grid. Then, a technique is used to couple those problems, using e.g. only some of the degrees of freedom on the boundaries to generate a global problem to be solved \cite{guiraldello2018multiscale}. 

The Multiscale Sampling Method follows roughly this same multiscale framework, and this is highlighted in Table~\ref{table:comparison}. The remaining of this subsection will show how the steps in this framework were adapted into a Markov Chain Monte Carlo setting, resulting in our new method. 

\begin{table}[h!]
\caption{How the Multiscale Sampling Method was motivated by the multiscale finite element methods}
\label{table:comparison}
\begin{tabular}{l p{100mm}}
\hline
\multicolumn{2}{l}{Multiscale PDE Solver $\leftrightarrow$ Multiscale Sampling Method}\\
\hline
Step A. & Partition $\Omega$ into non-overlapping subdomains\\
Step B. & Construct local basis functions\\
Step C. & Simulate on each subdomain separately\\
Step D. & Enforce coupling conditions between subdomains and solve the global problem\\
\hline
\end{tabular}
\end{table}

\begin{figure}[h!]
\centering
\includegraphics{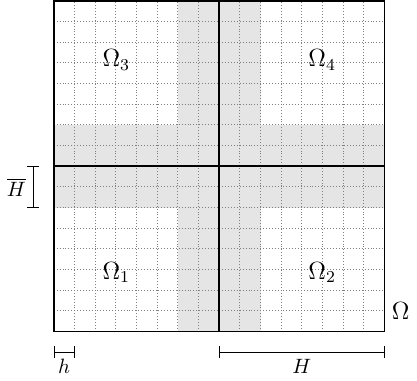}
\caption{Example of a domain $\Omega$ partitioned into $M_C=4$ subdomains $\Omega_i$ and their respective scales $h$, $H$, and $\overline{H}$. In gray are the cells in which a coupling procedure will be performed} 
\label{fig:scales}
\end{figure}

\vspace{0.5cm}
The name ``multiscale'' suggests that there are distinct scales in the method.  They are described in Table \ref{table:MSMscales} and depicted in Figure \ref{fig:scales}. A good choice for those parameters depends on the correlation lengths of the prior (see end of Section \ref{sec:inverseproblem}).

\begin{table}[h!]
\caption{Description of the three scales in the Multiscale Sampling Method and some notation on dimensions}
\label{table:MSMscales}
\begin{tabular}{l p{107mm}}
\hline
$h$ & This is the scale of the fine grid that solves the PDE\\
$H$ & This is the scale that partitions the domain $\Omega$ into non-overlapping, weakly coupled subdomains $\Omega_i$\\
$\overline{H}$ & This scale controls the band close to subdomain boundaries in which a coupling strategy is performed\\
\hline
$N_C$ & Number of terms in the local Karhunen-Lo\`eve expansion\\
$M_C$ & Number of subdomains\\
$N$   & Total stochastic dimension of the problem, $N=M_C*N_C$\\
\hline
\end{tabular}
\end{table}

\begin{figure}[h!]
\centering
\includegraphics{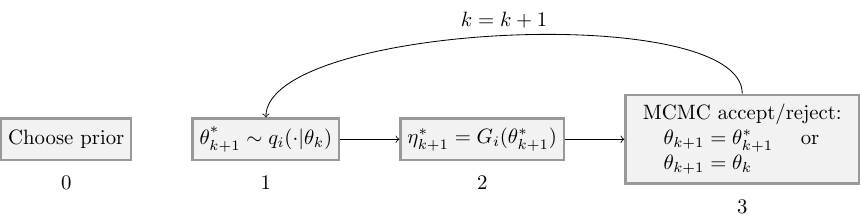}
\caption{Diagram of the states of our MCMC approach. Without coupling, our method is simply a Two Stage Delayed Acceptance-within-Gibbs with a different prior distribution. The coupling procedure in State 2 complicates the analysis of method}
\label{fig:MCMCdiagram}
\end{figure}

Let us see how the multiscale framework from Table \ref{table:comparison} was applied to construct the Multiscale Sampling Method, using all of the Markov Chain Monte Carlo ideas in Section \ref{sec:background}. Figure \ref{fig:MCMCdiagram} is a diagram showing a typical loop for the simulation of one value of a Markov chain. Each of those steps in the diagram will be referred to as a \textit{state}, leaving the term \textit{step} for when we mention a step of the multiscale framework from Table \ref{table:comparison}.  

\vspace{0.5cm}
\noindent\textbf{State 0:}

First, we decide the prior distribution for the field to be simulated. In line with \textit{Step A} from the multiscale framework (Table \ref{table:comparison}), the domain of interest $\Omega$ is partitioned into $M_C$ non-overlapping subdomains $\Omega_i$. The size of $\Omega_i$ is controlled by the scale $H$ (see Figure \ref{fig:scales} or Table \ref{table:MSMscales}). 

Now, as in \textit{Step B} of the multiscale framework, each subdomain $\Omega_i$ will have its own local prior: a Gaussian field with the same covariance structure as the global problem. A Karhunen-Lo\`eve Expansion (Section \ref{sec:KLE}), truncated at $N_C$ terms, gives us the $N_C$ local basis functions $\{\psi_{i1}, \psi_{i2}, \ldots \psi_{iN_C}\}$ and vector of parameters $\thetab_i:=(\theta_{i1},\theta_{i_2},\ldots, \theta_{iN_c})$ with prior $\pi_{i,\thetab}\sim N(0, Id_{N_C})$. 

Putting all the coefficients together into one big vector, we have 
\[\thetab := (\thetab_1,\ldots,\thetab_{M_C}), \qquad\pi_{\thetab}\sim N(0, Id_N).\]
This global prior considers the subdomains $\Omega_i$ to be independent from each other. 

\vspace{0.5cm}
\noindent\textbf{State 1:}

Moving on to the simulation loop, the first task is to generate a new sample value. As indicated by \textit{Step C} in the multiscale framework, we aim is to do this as locally as possible. This is where the Gibbs Sampler approach becomes useful. At each iteration of the MCMC, we will modify only one subdomain $\Omega_i$ at a time. Since the conditional posterior for the Gibbs sampler is not easily available for our class of inverse problems, we resort to a Metropolis-within-Gibbs idea. 

In principle, we can choose any instrumental distribution $q_i$, such as the Random Walk Sampler or preconditioned Crank-Nicolson (Section \ref{sec:pCN}). What matters is that these $q_i$ only change the coefficients of $\thetab$ that are inside the current subdomain of interest $\Omega_i$, and leave all other coefficients frozen.  

\vspace{0.5cm}
\noindent\textbf{State 2:}

As in \textit{Step D} in Table \ref{table:comparison}, once a new sample value $\thetab^*$ is generated, we must assemble the global field of interest, $\eta$, before evaluating the likelihood function for the MCMC. In non-multiscale approaches where the Karhunen-Lo\`eve decomposition is global, this is a trivial step (see discussion in Section \ref{sec:KLE}). But in the case of the Multiscale Sampling Methods, this step complicates the analysis of the new MCMC: the challenge lies in understanding how the global assembling operator $G$ will modify the state space of the MCMC and maybe even the instrumental distribution. 

Moreover, $G$ can either be independent of the active subspace $\Omega_i$, or depend on it. That's why the diagram and algorithm have $G$ with subscript $i$. In the next section, three different assembling operations $G$ will be studied. For example, the region inside $\overline{H}$ (see Figure \ref{fig:scales}) can be averaged in a certain way so as to couple the sample of the subdomains.  

\vspace{0.5cm}
\noindent\textbf{State 3:}

Using the global $\eta^*$, a MCMC approach is chosen to accept or reject the new value. This can be either a Metropolis-Hastings or Two-Stage Delayed-Acceptance method. The care in this state is to ensure that the probability of acceptance $\alpha$ given $q_i$ is correct. 

\vspace{0.5cm}
A quick summary of the above discussion can be found in Table \ref{table:MSMwithMCMC}. 

\begin{table}[h!]
\caption{How existing MCMC techniques fit the Multiscale Sampling Method}
\label{table:MSMwithMCMC}
\begin{tabular}{l p{102mm}}
\hline
State 0 & Decide the prior distribution: decompose the domain $\Omega$ into subdomains and use a Karhunen-Lo\`eve expansion\\
State 1 & Use a Gibbs approach to decide which components of $\thetab$ to update and whether to use a Random Walk Sampler or preconditioned Crank-Nicolson\\
State 2 & Construct the global log-permeability field $\etab$ field using $\thetab$. Here's where the coupling takes place\\
State 3 & Use either a Two-Stage Delayed Acceptance or Metropolis-Hastings step to accept the new sample or repeat the previously accepted one\\
\hline
\end{tabular}
\end{table}

Finally, Algorithm \ref{alg:msm} shows a Metropolis-within-Gibbs cycle of the Multiscale Sampling Method, using a Two-Stage Delayed-Acceptance method. It is adapted from the original algorithm in \cite{ali2024multiscale}.

\begin{algorithm}
\caption{Multiscale Sampling Method}
\label{alg:msm}
\begin{algorithmic}[1]
\STATE{Given $\thetab^{(t)}$}
\FOR{$i=1$ to $M_C$}
\STATE{Draw $\thetab^*\sim q_i(\cdot|\thetab^{(t)})$}
\STATE{Assemble the global field $\eta^*=G(\thetab^*)$}
\STATE{Evaluate the coarse approximation to the likelihood, $L_c(\eta^*)$}
\STATE{Promote $\thetab^*$ with probability $g(\thetab^{(t)}, \thetab^*)$,
\[g(x,y)=\min\left\{1,\frac{L_c(G(y))\pi_{\thetab}(y)}{L_c(G(x))\pi_{\thetab}(x)}\frac{q_i(x|y)}{q_i(y|x)}\right\}, \] otherwise set $\thetab^{(t+1)}=\thetab^{(t)}$ and start a new iteration at Step 1}
\STATE{Evaluate the fine approximation $L_f(\eta^*)$}
\STATE{Set $\thetab^{(t+1)}=\thetab^*$ with probability $\rho(\thetab^{(t)}, \thetab^*)$, and $\thetab^{(t+1)}=\thetab^{(t)}$ otherwise. Here, \[\rho(x,y) =\min\left\{1,\frac{L_c(G(x))L_f(G(y))}{L_c(G(y))L_f(G(x))}\right\}.\]}
\ENDFOR
\end{algorithmic}
\end{algorithm}

%%%%%%%%%%%%%%%%%%%%%%%%%%%%%%%%%%%%%%%%%%%%%%%%%%%%%%%%%%%%%%%%%%%%%%%%%%%%%%%%%%%%%%%%%%%%%%
%%%%%%%%%%%%%%%%%%%%%%%%%%%%%%%%%%%%%%%%%%%%%%%%%%%%%%%%%%%%%%%%%%%%%%%%%%%%%%%%%%%%%%%%%%%%%%%%
\section{A new framework to analyze Mutiscale Sampling Methods}
\label{sec:framework}

In what follows, we propose a framework consisting of a series of questions to guide the analysis of a Multiscale Sampling Method. Some of those methods could be analyzed entirely in terms of $\thetab$ and its target posterior $p_{\thetab}$,
\[p_{\thetab}(\thetab) \propto L(G(\thetab))\pi_{\thetab}(\thetab).\]
However, as we mentioned in Section \ref{sec:KLE}, one of the reasons why we're proposing this new framework is to better compare multiscale methods and the approximations they're using for the space of functions for the fields $\etab$. 

For example, consider two distinct methods that start with the same construction for the vector of parameters $\thetab$. Assign to both the same prior $\pi_{\thetab}$, but with a different coupling strategy codified in the operator $G$. Denote by $\widehat{G}$ the operator of first method and $\widetilde{G}$ the one for the second method. The analysis that relies solely on $\thetab$ gives us the two distinct posterior distributions proportional to
\[L(\widehat{G}(\thetab))\pi_{\thetab}(\thetab),\qquad\text{vs.}\qquad L(\widetilde{G}(\thetab))\pi_{\thetab}(\thetab),\]
and very little insight is gained on what those different coupling strategies do to the resulting simulation. 

Moreover, it will be shown in Section \ref{sec:MSMoriginal} that our new approach makes it clear that there are structural issues with a certain multiscale method. This wouldn't be easy to detect by looking at the chain in $\thetab$ alone. 

The first part of our analysis framework is the novel part. We propose to make explicit the equivalent space of functions that are approximated by the Multiscale Sampling Method. This allows for an easy comparison between methods and checking how the coupling operator $G$ affects the way $L^2(\Omega)$ is approximated. This strategy can also be understood as an identification of the equivalent state space of the Markov chain, if it is seen as a chain on the fields $\etab$ rather than on $\thetab$. 

The second part is about the conditions for the convergence of the MCMC. Those were outlined in Section \ref{sec:background} as seven conditions, written in terms of the target distribution $f$ and the instrumental distribution $q$. This part of the analysis can be seen as just the traditional way of studying an MCMC method. 

\vspace{0.5cm}
\noindent\textbf{Part 1. Description of what's being simulated}

\vspace{0.2cm} 
\noindent\textbf{1.} \textit{What is the equivalent state space $V$ for $\etab$?}

The idea is to start with the functions in each subdomain $\Omega_i$ that can be represented by the dimensional reduction technique like the Karhunen-Lo\`eve expansion (see Section \ref{sec:KLE}). Our goal will be to find $V=Im(G)$, where $G$ is the global assembling operator that constructs the global field. For that, we apply $G$ until a pattern is identified. 

\vspace{0.2cm} 
\noindent\textbf{2.} \textit{What is the prior distribution on this set V?}

In the simplest case, the prior distribution of a new sample value for $\etab$ is exactly the distribution of the coefficients of the linear combination (Section \ref{sec:KLE}). If not, then it is possible to detect that the MCMC is not well defined in this step. 

\vspace{0.2cm} 
\noindent\textbf{3.} \textit{Describe the posterior distribution of $\etab$}

In this paper, this will be done by finding the expression for the first and second moments of $\etab$ (see \textit{Approach B} in Section \ref{sec:KLE}).

\vspace{0.5cm}
\noindent\textbf{Part 2. Conditions for convergence of the MCMC}

This part is straighforward if $\thetab$ can still be used as the underlying Markov chain and a preconditioned Crank-Nicolson or Random Walk Sampler is used without further changes (Section \ref{sec:pCN}). However, more sophisticated Multiscale Sampling Methods in the future might require analyzing the chain entirely from the point of view of $\etab$ instead, and all of the following items will have to be checked in detail.  

\vspace{0.2cm} 
\noindent\textbf{1.} \textit{What is the instrumental distribution $q_i$ on $V$?}

If possible, this will be done by inheriting the probability density $q_i(\cdot|\thetab)$. 

\vspace{0.2cm} 
\noindent\textbf{2.} \textit{How to calculate the probability density $q_i$?}

If no further transformation on $\thetab^*$ is used and the prior on $\thetab$ is still Gaussian, we can keep the expression from Section \ref{sec:pCN}. 

\vspace{0.2cm} 
\noindent\textbf{3.} \textit{Check Conditions 1-7}

Below are the conditions identified in our review of MCMC methods in Section \ref{sec:background}.  

\vspace{0.2cm}
\noindent\textbf{Condition 1.} The set $\text{supp}\ f$ is connected.

\vspace{0.2cm}
\noindent\textbf{Condition 2.} The support of $q_i$ contains the support of $f$, 
\[\bigcup_{x\in\text{supp}\ f}\text{supp}\ q_i(\cdot|x)\supset \text{supp}\ f.\]

\vspace{0.2cm}
\noindent\textbf{Condition 3.} \textit{Positivity} property on $q$:
\[q_i(y|x) > 0 \text{ for every }(x,y)\in\text{supp}\ f\times\text{supp}\ f.\]

\vspace{0.2cm}
\noindent\textbf{Condition 3'.} The function $f$ must be bounded and positive on every compact set of its support, and there exist positive numbers $\epsilon$ and $\delta$ such that \[q(y|x)>\epsilon\quad \text{if }|x-y|<\delta.\] 

\vspace{0.2cm}
\noindent\textbf{Condition 4.} There is a nonzero probability that a step will be rejected.

\vspace{0.2cm}
\noindent\textbf{Condition 5.} $q_i(y|x)>0$ implies $f_c(y)>0$,

\vspace{0.2cm}
\noindent\textbf{Condition 6'.} For any $x$ and $y$ in the state space of the Markov chain, 
\[q_i(y|x)>0\iff q_i(x|y)>0.\]
This is trivially satisfied if \textit{Condition 3} is true for $q_i$.

\vspace{0.2cm}
\noindent\textbf{Condition 7.} The distribution $f$ is such that its $r$-dimensional integral has finite Lebesgue integral over every $r$-dimensional coordinate hyperplane of the state space, for all $1\leq r\leq D$. Here, $D$ is the dimension of the vector of parameters.

\vspace{0.2cm}
To check those conditions, we focus the analysis on the new support for the posterior $f$. If this support is still connected, then the use of a preconditioned Crank-Nicolson step or a Random Walk Sampler for instrumental distribution $q_i$ are enough to guarantee \textit{Conditions 1, 2, 3 and 6'}. This is because their step is a Gaussian step that has positive property to reach any other point in the state space (see Section \ref{sec:pCN}). 

For example, if we can still consider the Markov chain to be run on $\thetab$, then by 
\[f(\thetab) = \frac{L(G(\thetab))\pi_{\thetab}(\thetab)}{\int_{\mathcal{R}^N}L(G(\mathbf{u}))\pi_{\thetab}(\mathbf{u})d\mathbf{u}},\]
we see that the support of $f$ is the whole state space of $\thetab$, since the likelihood $L$ is an exponential function, and so is the prior $\pi_{\thetab}$.

\textit{Condition 4} is usually assumed in practice \cite{casella2002statistical}, and \textit{Condition 5} depends on $f_c(y)>0$ being positive in all its support set, which can be assumed here since the likelihood function comes from a Gaussian. 

Finally, \textit{Condition 7} will be left as an open problem for all methods described in the next section.

%%%%%%%%%%%%%%%%%%%%%%%%%%%%%%%%%%%%%%%%%%%%%%%%%%%%%%%%%%%%%%%%%%%%%%%%%%%%%%%%%%%%%%%
%%%%%%%%%%%%%%%%%%%%%%%%%%%%%%%%%%%%%%%%%%%%%%%%%%%%%%%%%%%%%%%%%%%%%%%%%%%%%%%%%%%%%%%
\section{Analysis of some multiscale sampling methods}
\label{sec:analysis}

We now showcase how the new framework to analyze multiscale sampling methods can be applied.
Three examples of multiscale methods will be presented and studied in increasing order of complexity, culminating in the original multiscale method with local averaging from \cite{ali2024multiscale}. 

All methods in this section have the same initial construction given in \textit{State 0} from Section \ref{sec:multiscale}, which will be recalled here. For simplicity, consider a square domain $\Omega$ partitioned into square subdomains $\Omega_i$ of length $H$. An underlying probability space is denoted by $\mathcal{X}=(\mathcal{S}, \boldsymbol{\sigma}(\mathcal{S}), \mathcal{P})$.  

Start with independent Gaussian fields $\etab_i\colon(\Omega_i, \mathcal{X})\rightarrow\mathcal{R}$ in each $\Omega_i$, and apply a Karhunen-Lo\`eve decomposition on them. Then, each $\Omega_i$ has its own basis of functions $\{\psi_{i1}, \psi_{i2}, \ldots, \psi_{iN_C}\}$. That is, before any coupling procedure is applied, the local fields inside each $\Omega_i$ look like
\[\etab_i(x, \omega) = \sum_{j=1}^{N_C}\sqrt{\lambda_{ij}}\theta_{ij}(\omega)\psi_{ij}(x), \qquad x\in\Omega_i,\ \omega\in\mathcal{S}.\]

For the rest of the section, the probability space $\mathcal{S}$ will be omitted, since it won't play any role in the discussion. The complete vector of parameters to be simulated is the concatenation of the parameters of each subdomain,
\[\thetab = (\theta_{11}, \ldots, \theta_{1N_C}, \theta_{21}, \ldots, \theta_{M_CN_C}).\]

The multiscale methods will now differ only in how the local fields are coupled together by the operator $G\colon \thetab\mapsto \etab$.

%%%%%%%%%%%%%%%%%%%%%%%%%%%%%%%%%%%%%%%%%%%%%%%%%%%%%%%%%%%%%%%%%%%%%%%%%%%%%%%%%%
\subsection{Multiscale sampling with no coupling}
\label{sec:MSMuncoupled}

The first approach to multiscale sampling is to glue all those initial $\etab_i$ fields on the subdomains without coupling them in any way. That is, our operator $G$ satisfies, inside any $\Omega_i$,
\[G(\thetab)(x) = \etab_i(x) = \sum_{j=1}^{N_C}\sqrt{\lambda_{ij}}\theta_{ij}\psi_{ij}(x), \qquad x\in\Omega_i.\]
This method has recently appeared as the \textit{stitched method} in \cite{xu2024domain}, where those authors show how to find a global basis of functions by trivial extensions of the local basis functions. However, that result is the only overlap between our work and theirs, and the rest of our discussion is new.

\subsubsection{Description of the equivalent state space $V$}

\vspace{0.2cm}
\noindent\textit{1. What is the equivalent state space for $\etab$?}

We'll show that, by using a trivial extension of the basis functions $\{\psi_{ij}\}_{ij}$ to the global domain $\Omega$, any new field generated by the method and coupled by $G$ will be a linear combination of $N$ global basis functions. This idea has already been published in \cite{xu2024domain}, with different notations and exposition. 

Extend each basis function $\psi_{ij}\colon\Omega_i\rightarrow\mathcal{R}$ to the whole domain, and denote it by $\Psi_{ij}\colon\Omega\rightarrow\mathcal{R}$. The extensions satisfiy
\[\Psi_{ij}(x) = \left\{\begin{array}{l l} \psi_{ij}(x), & \text{if }x\in\Omega_i\\
                        0, &\text{otherwise.}\end{array}\right.\] 
Then, we can write
\[\etab(x)=G(\thetab)(x) = \sum_{i=1}^{M_C}\sum_{j=1}^{N_C}\sqrt{\lambda_{ij}}\theta_{ij}\Psi_{ij}(x),\qquad\forall x\in\Omega.\]

Rename the indices for $\thetab$ as $\thetab=(\theta_1,\theta_2,\ldots, \theta_N)$ and analogously for $\{\Psi_{ij}\}_{i,j}$ and $\lambda_{ij}$. The expression for the field $\etab$ simplifies to
\[\etab(x)=G(\thetab)(x) = \sum_{j=1}^N\sqrt{\lambda_j}\theta_j\Psi_j(x),\qquad\forall x\in\Omega.\]

Hence, the equivalent state space for $\etab$ is 
\[V = \text{span}\{\Psi_1, \Psi_2, \ldots, \Psi_N\}= \text{span}\{\Psi_{ij}\}_{ij}.\]

\vspace{0.2cm}
\noindent\textit{2. What is the prior distribution on this set $V$?}

We follow the ideas in Section \ref{sec:KLE}. A function $\etab\in V$ has the form
\[\etab = \sum_{i=1}^Na_i\Psi_i,\]
so its equivalent prior probability is 
\[\pi_{\etab}(\etab)=\pi_{\thetab}\left(\frac{a_1}{\sqrt{\lambda_1}}, \frac{a_2}{\sqrt{\lambda_2}},\ldots,\frac{a_N}{\sqrt{\lambda_N}}\right).\]

\vspace{0.2cm}
\noindent\textit{3. How can we describe the posterior?}

The mean on each point in the field is just the average of the posterior of the entries of $\thetab$ used at that subdomain,
\[E_{p_{\etab}}[\etab(x)] = \sum_{k=1}^{N_C}\sqrt{\lambda_{ik}}\psi_{ik}(x)E_{p_{\theta_{ik}}}[\theta_{ij}],\qquad x\in\Omega_i.\]

For the covariance structure, we consider two cases. First, assuming $x_1\in\Omega_i$, $x_2\in \Omega_j$, and $i\neq j$, then

\begin{align*}
\Cov_{p_{\etab}}(\etab(x_1),\etab(x_2)) &= 
      \Cov_{p_{\boldsymbol{\theta}}}\left(\sum_{k=1}^{N_C}\sqrt{\lambda_{ik}}\psi_{ik}(x_1)\theta_{ik},
      \sum_{l=1}^{N_C}\sqrt{\lambda_{jl}}\psi_{jl}(x_2)\theta_{jl}\right) \\
   &= \sum_{k=1}^{N_C}\sum_{l=1}^{N_C}\sqrt{\lambda_{ik}\lambda_{jl}}
      \psi_{ik}(x_1)\psi_{jl}(x_2)\Cov_{p_{\boldsymbol{\theta}}}(\theta_{ik},\theta_{jl}).
\end{align*}
By our construction, the prior on $\thetab$ was such that each parameter was independent. This also makes the field restricted to each subdomain $\Omega_i$ to be independent from the other subdomains. On the posterior, a correlation between the entries of $\thetab$ will exist, and they are given by how the likelihood will connect them together during the acceptance or rejection in the Markov chain. This correlation will be present in the posterior of the fields as well, and it is given by the formula above.   

Something similar happens when $i=j$, and both points are in the same subdomain,
\begin{align*}
\Cov_{p_{\etab}}(\etab(x_1),\etab(x_2))= &
  \sum_{k=1}^{N_C}\lambda_{ik}\psi_{ik}(x_1)\psi_{ik}(x_2)\Var_{p_{\boldsymbol{\theta}}}[\theta_{ik}]\\
  &+2\sum_{k>l}\sqrt{\lambda_{ik}\lambda_{il}}\psi_{ik}(x_1)\psi_{il}(x_2)\Cov_{p_{\boldsymbol{\theta}}}(\theta_{ik},\theta_{il}).
\end{align*}
Here, there are two contributions to the covariance: the first sum was expected, because the same basis functions are used for both points inside the same subdomain $\Omega_i$. The second sum is given by the new covariance on the posterior distribution of $\thetab$. 

\subsubsection{Convergence conditions for the MCMC}

\vspace{0.2cm}
\noindent\textit{1. What is the instrumental distribution $q_i$ on $V$?}

Because of the one-to-one correspondence between $\etab$ and $\thetab$, the probability $q(\etab^*|\etab)$ is the same as that of its underlying parameters, $q(\thetab^*|\thetab)$. Both the preconditioned Crank-Nicolson or Random Walk Sampler from Section \ref{sec:pCN} can work as $q$.  

\vspace{0.2cm}
\noindent\textit{2. How to calculate the probability density $q_i$?}

Since no further transformation is done on the sampling of $\thetab$, the probability density of the instrumental distribution can be calculated by the formulas in Section \ref{sec:pCN}. 

\vspace{0.2cm}
\noindent\textit{3. Check Conditions 1-7}

This is the sample case mentioned in the presentation of of our framework. The chain runs on $\thetab$, so both the preconditined Crank-Nicolson and Random Walk Sampler can reach any function in $V$. The state space for $\thetab$ is still $\mathcal{R}^N$, so the support set of the posterior distribution is connected. Therefore, apart from \textit{Condition 7}, all conditions for convergence are satisfied.

\subsubsection{Discussion}

In summary, the equivalent state space of random fields for this method is a glued patch of local fields, which are only mixed by the likelihood function during the MCMC procedure. Still, we may want the sample fields to have better continuity properties between subdomains. For this, the next multiscale methods will have the operator $G$ to perform some averaging procedure on the new field.

%%%%%%%%%%%%%%%%%%%%%%%%%%%%%%%%%%%%%%%%%%%%%%%%%%%%%%%%%%%%%%%%%%%%%%%%%%%%%%%%%%%%%%%
\subsection{Multiscale sampling with global averaging}
\label{sec:MSMglobal}

In this method, the operator $G$ will average the value of $\etab_i$ at any point in the discretized grid that is within a distance $\overline{H}$ to a common boundary between subdomains. Those are the cells in gray in Figure \ref{fig:scales} from Section \ref{sec:multiscale}. The averaging procedure will use all points within an ellipse centered at the point of interest whose semi-axes have length that is a proportion of the correlation lengths $L_x$ and $L_y$ of the prior model (see Section \ref{sec:inverseproblem}, or \cite{ali2024multiscale}). 

The term \textit{global} was added to the name of this method because the averaging will be applied to every subdomain $\Omega_i$ at each iteration. This is opposed to the original Multiscale Sampling Method in \cite{ali2024multiscale}, that we call here \textit{locally averaged}. It performs the averaging only inside the active subdomain for the current iteration. 

\subsubsection{Description of the equivalent state space $V$}

Our analysis will be performed on two simple one-dimensional examples, and some comments on how to extend this technique to two dimensions are given at the end. Also for simplicity, this analysis will depend on the discretization of the grid, and its generalization to a more continuous form is left for a future project. 

\begin{figure}[h!]
\centering
\includegraphics{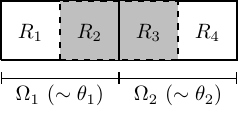}
\caption{Domain for the examples. There are two non-overlapping subdomains $\Omega_1$ and $\Omega_2$, and an averaging procedure will be applied to values of $\etab$ inside the intermediate regions $R_2$ and $R_3$. }
\label{fig:averagingex}
\end{figure}

Consider the one-dimensional set $\Omega=\Omega_1\cup\Omega_2$ in Figure~\ref{fig:averagingex}. Each subdomain $\Omega_i$ has just one basis function $\psi_i$, a corresponding eigenvalue $\lambda_i$, and a parameter $\theta_i$. Depicted in gray are the subsets $R_2\subset \Omega_1$ and $R_3\subset \Omega_2$  which are close to the common boundary $\partial\Omega_1\cap\partial\Omega_2$ and on which averaging will be done.

Before applying $G$, a sampled field $\widetilde{\etab}$ with parameter vector $\thetab=(\theta_1,\theta_2)$ looks like
\[\widetilde{\etab}(x) = \left\{\begin{array}{ll}
         \sqrt{\lambda_1}\theta_1\psi_1(x),& x\in\Omega_1,\\[0.2cm]
         \sqrt{\lambda_2}\theta_2\psi_2(x),& x\in\Omega_2.\\
         \end{array}\right.\]

\vspace{0.5cm}
\noindent\textbf{Example A.} Suppose that $R_2$ and $R_3$ have only one point each in the discretized domain, and that averaging is performed such that the value in $R_2$ is only averaged with the value of the point in $R_3$, and the analogous happens for the point in $R_3$. For the purposes of averaging, the values at each of those points is that of the uncoupled field $\widetilde{\etab}$ above. 

In agreement with the different scales of a multiscale method, the letter $h$ denotes the length on the grid that takes us to the next point (see Figure \ref{fig:scales}). Regions $R_1$ and $R_4$ will not be affected by the coupling, and the coupled field $\etab$ reads

\[\etab(x) = G(\thetab)(x) := \left\{\begin{array}{ll}
                   \sqrt{\lambda_1}\psi_1(x)\theta_1, & x\in R_1,\\[0.2cm]
                   \frac{1}{2}\left(\sqrt{\lambda_1}\psi_1(x)\theta_1+\sqrt{\lambda_2}\psi_2(x+h)\theta_2\right), &  x\in R_2,\\[0.2cm]
                   \frac{1}{2}\left(\sqrt{\lambda_1}\psi_1(x-h)\theta_1+\sqrt{\lambda_2}\psi_2(x)\theta_2\right), & x\in R_3,\\[0.2cm]
                   \sqrt{\lambda_2}\psi_2(x)\theta_2, & x\in R_4.
                   \end{array}\right.\]

In the multiscale MCMC algorithm, a new sample field will be drawn by choosing either $\Omega_1$ and $\Omega_2$ to be \textit{active} for that iteration. A new corresponding $\theta_i$ is drawn using the instrumental distribution $q_i$, while the other $\theta$ remains fixed as the previous one. Then, averaging is performed on the new field to yield $\etab$. For example, if at iteration $t$ the active subdomain is $\Omega_1$, then a new $\theta_1$ is simulated. Although $\theta_2$ will be the same as in iteration $t-1$, the value of the global field will still change in $R_3$. The multiscale method in the next section will keep the value of $\etab$ in $R_3$ unchanged. 

We can find a global basis for the fields $\etab$, as in the uncoupled method of the previous section. Based on $\psi_i\colon\Omega_1\rightarrow\mathcal{R}$, the global function $\Psi_1\colon\Omega\rightarrow\mathcal{R}$ is written the following way,
\[\Psi_1(x) = \left\{\begin{array}{ll}
                   \psi_1(x), & x\in R_1,\\[0.2cm]
                   \frac{1}{2}\psi_1(x),& x\in R_2,\\[0.2cm]
                   \frac{1}{2}\psi_1(x-h),& x\in R_3,\\[0.2cm]
                   0, & x\in R_4.
                   \end{array}\right.\]
Analogously, $\Psi_2\colon\Omega\rightarrow\mathcal{R}$ is based on $\psi_2\colon\Omega_2\rightarrow\mathcal{R}$:
\[\Psi_2(x) = \left\{\begin{array}{ll}
                   0, & x\in R_1,\\[0.2cm]
                   \frac{1}{2}\psi_2(x+h),&x\in R_2,\\[0.2cm]
                   \frac{1}{2}\psi_2(x),&x\in R_3,\\[0.2cm]
                   \psi_2(x),&x\in R_4.\\
                   \end{array}\right.\]
                   
Then, any $\etab$ sampled by this method is in the space
\[V = \text{span}\{\Psi_1, \Psi_2\}.\]
   
Therefore, the construction used by this method depends only on the current proposal for $\thetab$, and the relationship between $\etab$ and $\thetab$ is one-to-one. This means that the analysis of the MCMC method can be done in terms of $\thetab$, and the probability structure for $\etab$ can be inherited from the one for $\thetab$.                    

\vspace{0.5cm}                   
\noindent\textbf{Example B.} The setting for this example is the same as before, but now the ellipse that chooses which points to use in the averaging of the points in grey is assumed big enough to involve the points in $R_1$ and $R_4$. That is, now the point in $R_2$ will be averaged with the point in $R_3$ and its neighboring point in $R_1$. Analogously, theh point in $R_3$ will be averaged with the one in $R_2$ and its immediate neighbor in $R_4$. 

The coupling operator $G$ can be written as 
\[\etab(x) = G(\thetab)(x) := \left\{\begin{array}{ll}
                   \sqrt{\lambda_1}\psi_1(x)\theta_1, & x\in R_1,\\[0.2cm]
                   \frac{1}{3}\bigg(\sqrt{\lambda_1}\big(\psi_1(x-h)+\psi_1(x)\big)\theta_1+\sqrt{\lambda_2}\psi_2(x+h)\theta_2\bigg), & x\in R_2,\\[0.5cm]
                   \frac{1}{3}\bigg(\sqrt{\lambda_1}\psi_1(x-h)\theta_1+\sqrt{\lambda_2}\big(\psi_2(x)+\psi_2(x+h)\big)\theta_2\bigg), & x\in R_3,\\[0.5cm]
                   \sqrt{\lambda_2}\psi_2(x)\theta_2, & x\in R_4.
                   \end{array}\right.\]

The new global basis functions are then $\Psi_1\colon\Omega\rightarrow\mathcal{R}$,
\[\Psi_1(x) = \left\{\begin{array}{ll}
                     \psi_1(x), & x\in R_1,\\[0.2cm]
                     \frac{1}{3}\left(\psi_1(x-h)+\psi_1(x)\right), & x \in R_2,\\[0.2cm]
                     \frac{1}{3}\psi_1(x-h), & x\in R_3,\\[0.2cm]
                     0, & x\in R_4,\\
                     \end{array}\right.\] 
and $\Psi_2\colon\Omega\rightarrow\mathcal{R}$,
\[\Psi_2(x) = \left\{\begin{array}{ll}
                     0, & x\in R_1,\\[0.2cm]
                     \frac{1}{3}\psi_2(x+h), & x\in R_2,\\[0.2cm]
                     \frac{1}{3}\left(\psi_2(x)+\psi_2(x+h)\right), & x \in R_3,\\[0.2cm]
                     \psi_2(x), & x\in R_4.
                     \end{array}\right.\]
Any $\etab$ sampled from the method is in 
\[V=\text{span}\{\Psi_1, \Psi_2\}.\]   

Notice that, by increasing the size of the ellipse which chooses how many points will participate in the averaging, we are increasing spatial correlations between the points in the gray regions $R_2$ and $R_3$ and the rest of their subdomains.   

\begin{figure}[h!]
\centering
\includegraphics{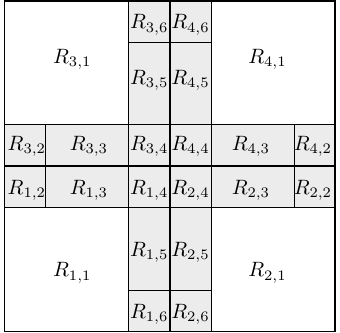}
\caption{In 2D, the construction of $V$ will require the analysis on the 24 subsets $R_{i,j}$}
\label{fig:average2D}
\end{figure}

It is straighforward to generalize those examples to more points and basis functions. They will follow the same symmetry as above, and each extension to the global basis function will be either like the case for $\Psi_1$ or $\Psi_2$. For the 2D problem with four subdomains $\Omega_i$, we would have to consider 24 distinct subsets as in the Figure \ref{fig:average2D}. This is because each corner will have a different formula for $G(x)$. 

We can now answer the questions in our framework.

\vspace{0.5cm}
\noindent\textit{1. What is the equivalent state space $V$ for $\etab$?} 

Those were constructed for each example above.

\vspace{0.5cm}
\noindent\textit{2. What is the prior distribution on this set $V$?} 

As in the previous method, the fact that there exists a one-to-one correspondence between $\etab$ and $\thetab$ means that we can calculate this prior exactly as in Section \ref{sec:MSMuncoupled}.

If $\etab=a\Psi_1+b\Psi_2$, then
\[\pi_{\etab}(\etab)=\pi_{\thetab}\left(\frac{a}{\sqrt{\lambda_1}},\frac{b}{\sqrt{\lambda_2}}\right).\]

\vspace{0.5cm}
\noindent\textit{3. How can we describe the posterior distribution on $V?$} 

The first two moments of the fields for \textit{Example A} will be expressed below in terms of the posterior for $\thetab$, and an analogous study can be done for \textit{Example B} in the future. These moments will be compared with the expressions for the moments of the fields for the uncoupled method. However, an important remark is that the covariance structure for the posterior of $\thetab$, $\Cov_{p_{\thetab}}(\cdot,\cdot)$, will be different for the globally averaged and uncoupled method. This means that this analysis does not yet tell the full story when it comes to comparing the fields simulated by different methods. 

At each point, the posterior mean can be calculated by
\[E_{p_{\etab}}[\etab(x)] = \sum_{l=1}^2\sqrt{\lambda_l}\Psi_l(x)E_{p_{\theta_l}}[\theta_l],\qquad x\in\Omega.\]

While in $R_1$ and $R_4$ the expected value of the field depends only on one parameter $\theta_i$, this will change for $R_2$ and $R_3$. For example, in $R_2$,
\[E_{p_{\etab}}[\etab(x)] = \frac{1}{2}\left(\sqrt{\lambda_1}\psi_1(x)E_{p_{\theta_1}}[\theta_1]+\sqrt{\lambda_2}\psi_2(x+h)E_{p_{\theta_2}}[\theta_2]\right)\]
The expresions in the sum are the mean of the field at the points in $R_2$ and $R_3$, respectively, for the uncoupled method of Section \ref{sec:MSMuncoupled}.

For the covariance structure, a general expression is
\begin{align*}
\Cov_{p_{\etab}}(\etab(x_1),\etab(x_2)) &= 
      \Cov_{p_{\boldsymbol{\theta}}}\left(\sum_{k=1}^{2}\sqrt{\lambda_k}\Psi_k(x_1)\theta_k,
      \sum_{l=1}^2\sqrt{\lambda_l}\Psi_l(x_2)\theta_l\right) \\
   &= \sum_{k=1}^2\sum_{l=1}^2\sqrt{\lambda_k\lambda_l}
      \Psi_k(x_1)\Psi_l(x_2)\Cov_{p_{\boldsymbol{\theta}}}(\theta_k,\theta_l),
\end{align*}
and the interesting cases are:

\vspace{0.2cm}
\noindent\textbf{Case 1.} $x_1\in R_1$, $x_2\in R_2$. 
\begin{align*}
\Cov_{p_{\etab}}&(\etab(x_1),\etab(x_2)) = 
      \lambda_1\Psi_1(x_1)\Psi_1(x_2)\Cov_{p_{\boldsymbol{\theta}}}(\theta_1,\theta_1)+
          \sqrt{\lambda_1\lambda_2}\Psi_1(x_1)\Psi_2(x_2)\Cov_{p_{\boldsymbol{\theta}}}(\theta_1,\theta_2)\\
      &= \frac{1}{2}\lambda_1\psi_1(x_1)\psi_1(x_2)\Var_{p_{\theta_1}}[\theta_1]+
          \frac{1}{2}\sqrt{\lambda_1\lambda_2}\psi_1(x_1)\psi_2(x_2+h)\Cov_{p_{\boldsymbol{\theta}}}(\theta_1,\theta_2)\\
\end{align*}
Compare it with the expression for the uncoupled method, 
\[\Cov_{p_{\etab}}(\etab(x_1),\etab(x_2)) = \lambda_1\psi_1(x_1)\psi_1(x_2)\Var_{p_{\theta_1}}[\theta_1].\]
The second term in the expression for the globally averaged method can be interpreted the following way. Even in the same subdomain $\Omega_1$, the covariance with a point close to a common boundary will incorporate the variation of the field in the adjacend subdomain. 

\vspace{0.2cm}
\noindent\textbf{Case 2.} $x_1\in R_1$, $x_2\in R_3$. 
\begin{align*}
\Cov_{p_{\etab}}&(\etab(x_1),\etab(x_2)) = 
      \lambda_1\Psi_1(x_1)\Psi_1(x_2)\Cov_{p_{\boldsymbol{\theta}}}(\theta_1,\theta_1)+
          \sqrt{\lambda_1\lambda_2}\Psi_1(x_1)\Psi_2(x_2)\Cov_{p_{\boldsymbol{\theta}}}(\theta_1,\theta_2)\\
      &= \frac{1}{2}\lambda_1\psi_1(x_1)\psi_1(x_2-h)\Var_{p_{\boldsymbol{\theta}}}[\theta_1]+
          \frac{1}{2}\sqrt{\lambda_1\lambda_2}\psi_1(x_1)\psi_2(x_2)\Cov_{p_{\boldsymbol{\theta}}}(\theta_1,\theta_2)\\
\end{align*}
Compare this with the expression for the uncoupled method, 
\[\Cov_{p_{\etab}}(\etab(x_1),\etab(x_2)) = \sqrt{\lambda_1\lambda_2}\psi_1(x_1)\psi_2(x_2)\Cov_{p_{\boldsymbol{\theta}}}(\theta_1,\theta_2).\]
In this case, the point $x_2-h$ is the point in $R_2$. Therefore, in the globally averaged method the covariance has an extra term, corresponding to the covariance of $x_1$ and $R_2$ from an uncoupled method. 

\vspace{0.2cm}
\noindent\textbf{Case 3.} $x_1\in R_2$, $x_2\in R_3$. 
\begin{align*}
\Cov_{p_{\etab}}&(\etab(x_1),\etab(x_2)) = 
      \sum_{k=1}^2\sum_{l=1}^2\sqrt{\lambda_k\lambda_l}
      \Psi_k(x_1)\Psi_l(x_2)\Cov_{p_{\boldsymbol{\theta}}}(\theta_k,\theta_l)\\
      &= \frac{1}{4}\lambda_1\psi_1(x_1)\psi_1(x_2-h)\Var_{p_{\boldsymbol{\theta}}}[\theta_1]+
          \frac{1}{4}\lambda_2\psi_2(x_1+h)\psi_2(x_2)\Var_{p_{\boldsymbol{\theta}}}[\theta_2]+\\
      &\qquad + \frac{1}{4}\sqrt{\lambda_1\lambda_2}\psi_1(x_1)\psi_2(x_2)\Cov_{p_{\boldsymbol{\theta}}}(\theta_1,\theta_2)\\
      &\qquad + \frac{1}{4}\sqrt{\lambda_2\lambda_1}\psi_1(x_1+h)\psi_2(x_2-h)\Cov_{p_{\boldsymbol{\theta}}}(\theta_2,\theta_1)\\
      &= \frac{1}{4}\lambda_1(\psi_1(x_1))^2\Var_{p_{\boldsymbol{\theta}}}[\theta_1]+
          \frac{1}{4}\lambda_2(\psi_2(x_2))^2\Var_{p_{\boldsymbol{\theta}}}[\theta_2]+\\
      &\qquad + \frac{1}{2}\sqrt{\lambda_1\lambda_2}\psi_1(x_1)\psi_2(x_2)\Cov_{p_{\boldsymbol{\theta}}}(\theta_1,\theta_2)\\
\end{align*}
The uncoupled formula for this covariance is
\[\Cov_{p_{\etab}}(\etab(x_1),\etab(x_2)) = \sqrt{\lambda_1\lambda_2}\psi_1(x_1)\psi_2(x_2)\Cov_{p_{\boldsymbol{\theta}}}(\theta_1,\theta_2).\]
That is, in the globally averaged method, the covariance between points in $R_2$ in $R_3$ has two extra terms that correspond to the variances in $R_1$ and $R_2$ from an uncoupled approach.  

\subsubsection{Convergence conditions for the MCMC}

Since each proposed $\etab$ has a one-to-one correspondence to a $\thetab$, the discussion for convergence is very similar to the one in the previous method.

\vspace{0.2cm}
\noindent\textit{1. What is the instrumental distribution $q_i$ on $V$?}

The instrumental distribution $q_i$ is written for $\etab$ via its corresponding parameter $\thetab$. That is, denoting by $\etab^*$ the new proposed field and by $\etab$ the last accepted field,   
 \[q_i(\etab^*|\etab) = q_i(\thetab^*|\thetab).\] 

If a Gibbs approach is used, then $q_i$ only modifies $\theta_i$. However, this method cannot fully isolate the modifications of $\etab$ on $\Omega_i$ only, since all adjacent subdomains will be affected. That is, it cannot be interpreted as Gibbs approach on the subdomains. 

\vspace{0.2cm}
\noindent\textit{2. How to calculate the probability density $q_i$?}

Since no transformation is applied to the new $\theta^*$, the density $q_i$ has the same formula as in Section \ref{sec:pCN}. 

\vspace{0.2cm}
\noindent\textit{3. Check Conditions 1-7}

The conditions for MCMC convergence follow because this is still a Markov chain on $\thetab$ that uses a Gaussian prior and a preconditioned Crank-Nicolson or random walk step. The one-to-one correspondence between $\thetab$ and $\etab$ also means that the instrumental distribution can generate any function in the space $V$ at each iteration. This will contrast with the next method to be analyzed.

%%%%%%%%%%%%%%%%%%%%%%%%%%%%%%%%%%%%%%%%%%%%%%%%%%%%%%%%%%%%%%%%%%%%%%%%%%%%%%%%%%%%%%%%%%%%
\subsection{Multiscale sampling with local averaging}
\label{sec:MSMoriginal}

In the previous method with global averaging, a new proposed sample field $\etab^*$ modifies the previous field $\eta$ in all adjacent subdomains to the active subdomain. The locally averaged method aims to keep all those adjacent subdomains fixed, and it was first proposed in \cite{ali2024multiscale}.

\subsubsection{Description of the equivalent state space $V$}

The analysis will be performed in an example that has the same setting as \textit{Example A} from Section \ref{sec:MSMglobal}. For convenience, its figure is repeated below. 

\begin{figure}[h!]
\centering
\includegraphics{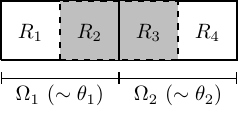}
\caption{Domain for the example of a multiscale method with local averaging. There are two non-overlapping subdomains $\Omega_1$ and $\Omega_2$, and an averaging procedure will be applied to values of $\etab$ inside the intermediate regions $R_2$ and $R_3$.}
\label{fig:averagingex2}
\end{figure}

Suppose that $R_2$ and $R_3$ contain only one point, and assume that only $R_2$ and $R_3$ are involved in the averaging. The analysis for the locally averaged method is more complicated: we must perform a few iterations of sampling and track what happens to the proposed fields. 

\vspace{0.2cm}
\noindent \textbf{Iteration 0.} Draw initial $\theta^0_1$ and $\theta^0_2$. The initial field $\etab^0$ is 
\[\eta^0(x) = \left\{\begin{array}{ll}
                     \sqrt{\lambda_1}\Psi_1(x)\theta_1^0, & x\in R_1\cup R_2,\\[0.2cm]
                     \sqrt{\lambda_1}\Psi_2(x)\theta_2^0, & x\in R_3\cup R_4,
                     \end{array}\right.\]
and it is accepted by default.    

\vspace{0.2cm}   
\noindent\textbf{Iteration 1. } The active subdomain is $\Omega_1$. Generate $\theta_1^1$ via a preconditioned Crank-Nicolson or random walk and perform averaging in $R_2$ only. The proposed new field $\etab^{1*}$ becomes
\[\etab^{1*}(x) = \left\{\begin{array}{ll}
                     \sqrt{\lambda_1}\Psi_1(x)\theta^1_1, & x\in R_1,\\[0.2cm]
                     \frac{1}{2}\left(\sqrt{\lambda_1}\Psi_1(x)\theta^1_1+\sqrt{\lambda_1}\Psi_2(x+h)\theta^0_2\right), & x\in R_2,\\[0.2cm]
                     \sqrt{\lambda_1}\Psi_2(x)\theta^0_2, & x\in R_3\cup R_4.
                     \end{array}\right.\]
   
Notice that the difference from this and the global averaging method is that now the field in $R_3$ is unchanged from the previous iteration. Also, $\theta_2$ will be repeated from the previous iteration, so $\theta^1_2=\theta_2^0$. To keep track of when it was last modified, we'll keep $\theta_2^0$ for the remainder of the example. 

If this step is accepted by the MCMC, then $\etab^1=\etab^{1*}$. Otherwise, $\etab^1=\etab^0$.    
   
\vspace{0.2cm}
\noindent\textbf{Iteration 2. } The active subdomain is $\Omega_2$. Generate a new value $\theta^2_2$ and perform averaging on $R_3$ using the previous $\etab^1$.

\[\etab^{2*}(x) = \left\{\begin{array}{ll}
                     \etab^1(x), & x\in R_1\cup R_2,\\[0.2cm]
                     \frac{1}{2}\left(\etab^1(x-h)+\sqrt{\lambda_1}\Psi_2(x)\theta^2_2\right), & x\in R_3,\\[0.2cm]
                     \sqrt{\lambda_1}\Psi_2(x)\theta^2_2, & x\in R_4.
                     \end{array}\right.\]

The extra complexity in analyzing this new method is that the construction depends on whether the previous proposed $\etab^*$ was accepted or not. Below, the expression for $\etab^{2*}$ in $R_3$ will be determined in each case, and the appropriate expressions for $R_1$ and $R_2$ are explicitly given. 

\vspace{0.5cm}
\noindent\textit{Case 1.} If $\etab^{1*}$ was accepted,      
\[\etab^{2*}(x) = \left\{\begin{array}{ll}
                     \sqrt{\lambda_1}\Psi_1(x)\theta^1_1, & x\in R_1,\\[0.2cm]
                     \frac{1}{2}\left(\sqrt{\lambda_1}\Psi_1(x)\theta^1_1+\sqrt{\lambda_1}\Psi_2(x+h)\theta^0_2\right), & x\in R_2,\\[0.2cm]
                     \frac{1}{2}\left(\etab^1(x-h)+\sqrt{\lambda_1}\Psi_2(x)\theta^2_2\right), & x\in R_3,\\[0.2cm]
                     \sqrt{\lambda_1}\Psi_2(x)\theta^2_2, & x\in R_4.
                     \end{array}\right.\]

In $R_3$, the expression becomes
\begin{align*}\etab^{2*}(x)&=\frac{1}{2}\left(\frac{1}{2}\left(\sqrt{\lambda_1}\Psi_1(x-h)\theta^1_1+\sqrt{\lambda_1}\Psi_2(x)\theta^0_2\right)+\sqrt{\lambda_1}\Psi_2(x)\theta^2_2\right)\\
                       &=\sqrt{\lambda_1}\Psi_1(x-h)\frac{1}{4}\theta_1^1+\sqrt{\lambda_1}\Psi_2(x)\left(\frac{1}{4}\theta_2^0+\frac{1}{2}\theta_2^2\right)
\end{align*}
   
\vspace{0.2cm}
\noindent\textit{Case 2.} If $\etab^{1*}$ was rejected   
 \[\etab^{2*}(x) = \left\{\begin{array}{ll}
                     \sqrt{\lambda_1}\Psi_1(x)\theta_1^0, & x\in R_1\cup R_2,\\[0.2cm]
                     \frac{1}{2}\left(\etab^1(x-h)+\sqrt{\lambda_1}\Psi_2(x)\theta^2_2\right), & x\in R_3,\\[0.2cm]
                     \sqrt{\lambda_1}\Psi_2(x)\theta^2_2, & x\in R_4.
                     \end{array}\right.\]  

And in $R_3$, 
\begin{align*}\etab^{2*}(x)&=\frac{1}{2}\left(\sqrt{\lambda_1}\Psi_1(x-h)\theta^0_1+\sqrt{\lambda_1}\Psi_2(x)\theta^2_2\right)\\
                       &=\sqrt{\lambda_1}\Psi_1(x-h)\frac{1}{2}\theta_1^0+\sqrt{\lambda_1}\Psi_2(x)\frac{1}{2}\theta_2^2
\end{align*}  
If the proposed new field is accepted, then $\etab^2=\etab^{2*}$. Otherwise, keep $\etab^2=\etab^1$.

\vspace{0.2cm}   
\noindent\textbf{Iteration 3.} The active subdomain is again $\Omega_1$. Generate a new $\theta^3_1$ and average on $R_2$ using the previous $\etab^2$.   
\[\etab^{3*}(x) = \left\{\begin{array}{ll}
                     \sqrt{\lambda_1}\Psi_1(x)\theta^3_1, & x\in R_1,\\[0.2cm]
                     \frac{1}{2}\left(\sqrt{\lambda_1}\Psi_1(x)\theta^3_1+\etab^2(x+h)\right), & x\in R_2,\\[0.2cm]
                     \etab^2(x), & x\in R_3\cup R_4.
                     \end{array}\right.\]

We must consider 4 cases. They correspond to the entire history of the chain so far. 

\vspace{0.5cm}
\noindent\textit{Case 1.} If all steps have been rejected so far, we have something like iteration 1:

\[\etab^{3*}(x) = \left\{\begin{array}{ll}
                     \sqrt{\lambda_1}\Psi_1(x)\theta^3_1, & x\in R_1,\\[0.2cm]
                     \frac{1}{2}\left(\sqrt{\lambda_1}\Psi_1(x)\theta^3_1+\sqrt{\lambda_1}\Psi_2(x+h)\theta^0_2\right), & x\in R_2,\\[0.2cm]
                     \sqrt{\lambda_1}\Psi_2(x)\theta^0_2, & x\in R_3\cup R_4.
                     \end{array}\right.\]

\vspace{0.5cm}
\noindent\textit{Case 2.} If the first proposed value was rejected but the second accepted:

\[\etab^{3*}(x) = \left\{\begin{array}{ll}
                     \sqrt{\lambda_1}\Psi_1(x)\theta^3_1, & x\in R_1,\\[0.2cm]
                     \frac{1}{2}\left(\sqrt{\lambda_1}\Psi_1(x)\theta^3_1+\etab^2(x+h)\right), & x\in R_2,\\[0.2cm]
                     \sqrt{\lambda_1}\Psi_1(x-h)\frac{1}{2}\theta_1^0+\sqrt{\lambda_1}\Psi_2(x)\frac{1}{2}\theta_2^2, & x\in R_3,\\[0.2cm]
                     \sqrt{\lambda_1}\Psi_2(x)\theta^2_2, & x\in R_4.
                     \end{array}\right.\]

And $R_2$ simplifies to
\begin{align*}\etab^3(x) &= \frac{1}{2}\left(\sqrt{\lambda_1}\Psi_1(x)\theta^3_1+\etab^2(x+h)\right) \\
 &=\frac{1}{2}\left(\sqrt{\lambda_1}\Psi_1(x)\theta^3_1+\sqrt{\lambda_1}\Psi_1(x)\frac{1}{2}\theta_1^0+\sqrt{\lambda_1}\Psi_2(x+h)\frac{1}{2}\theta_2^2\right) \\
 &=\sqrt{\lambda_1}\Psi_1(x)\left(\frac{1}{2}\theta^3_1+\frac{1}{4}\theta_1^0\right)+\sqrt{\lambda_1}\Psi_2(x+h)\frac{1}{4}\theta_2^2
\end{align*}

\vspace{0.2cm}
\noindent\textit{Case 3.} If the first proposed field was accepted but the second was rejected,

\[\etab^{3*}(x) = \left\{\begin{array}{ll}
                     \sqrt{\lambda_1}\Psi_1(x)\theta^3_1, & x\in R_1,\\[0.2cm]
                     \frac{1}{2}\left(\sqrt{\lambda_1}\Psi_1(x)\theta^3_1+\etab^1(x+h)\right), & x\in R_2,\\[0.2cm]
                     \sqrt{\lambda_1}\Psi_2(x)\theta^0_2, & x\in R_3\cup R_4.
                     \end{array}\right.\]   

In $R_2$, 
\begin{align*} \etab^3(x) &=\frac{1}{2}\left(\sqrt{\lambda_1}\Psi_1(x)\theta^3_1+\etab^1(x+h)\right)\\
&=\sqrt{\lambda_1}\Psi_1(x)\frac{1}{2}\theta^3_1+\sqrt{\lambda_1}\Psi_2(x+h)\frac{1}{2}\theta^0_2
\end{align*}

\vspace{0.2cm}
\noindent\textit{Case 4.} If all previous fields were accepted,       
\[\etab^3(x) = \left\{\begin{array}{ll}
                     \sqrt{\lambda_1}\Psi_1(x)\theta^3_1, & x\in R_1,\\[0.2cm]
                     \frac{1}{2}\left(\sqrt{\lambda_1}\Psi_1(x)\theta^3_1+\etab^2(x+h)\right), & x\in R_2,\\[0.2cm]
                      \sqrt{\lambda_1}\Psi_1(x-h)\frac{1}{4}\theta_1^1+\sqrt{\lambda_1}\Psi_2(x)\left(\frac{1}{4}\theta_2^0+\frac{1}{2}\theta_2^2\right)& x\in R_3,\\[0.2cm]
                     \sqrt{\lambda_1}\Psi_2(x)\theta^2_2, & x\in R_4.
                     \end{array}\right.\]
   
And in $R_2$,
\begin{align*}\etab^3(x)&=\frac{1}{2}\left(\sqrt{\lambda_1}\Psi_1(x)\theta^3_1+\sqrt{\lambda_1}\Psi_1(x)\frac{1}{4}\theta_1^1+\sqrt{\lambda_1}\Psi_2(x+h)\left(\frac{1}{4}\theta_2^0+\frac{1}{2}\theta_2^2\right)\right)\\
   &=\sqrt{\lambda_1}\Psi_1(x)\left(\frac{1}{2}\theta_1^3+\frac{1}{8}\theta_1^1\right)+\sqrt{\lambda_1}\Psi_2(x+h)\left(\frac{1}{8}\theta_2^0+\frac{1}{4}\theta_2^2\right)
\end{align*}

\begin{theorem}\label{theorem:msm} Suppose a proposal for each subdomain $\Omega_1$ and $\Omega_2$ was accepted at least once. Then there exist $a, b, c, d\in \mathcal{R}$ such that
\[\etab^*(x) =\left\{\begin{array}{ll}\sqrt{\lambda_1}\Psi_1(x)a+\sqrt{\lambda_2}\Psi_2(x+h)b, & x\in R_2,\\[0.2cm]
                    \sqrt{\lambda_1}\Psi_1(x-h)c+\sqrt{\lambda_2}\Psi_2(x)d,& x\in R_3.
                    \end{array}\right.\]
\end{theorem}
\begin{proof}
The proof is by induction on the accepted iterations. The base case is given in the example above. The first time a new value is accepted in $\Omega_1$, then the expression in $R_2$ becomes of the form in the theorem. Analogously, the first time a new value is accepted in $\Omega_2$, then the field in $R_3$ takes the desired form. 

Now, denote by $\etab$ the previously accepted value and that $\Omega_1$ is the active subdomain for this iteration $t$. The expression in $R_2$ for the new field is
\begin{align*}\etab^{(t)}(x) &= \frac{1}{2}\left(\sqrt{\lambda_1}\Psi_1(x)\theta^{(t)}_1+\etab(x+h)\right)\\
 &=\frac{1}{2}\left(\sqrt{\lambda_1}\Psi_1(x)\theta^{(t)}_1+\left(\sqrt{\lambda_1}\Psi_1(x)c+\sqrt{\lambda_2}\Psi_2(x+h)d\right)\right)\\
 &=\sqrt{\lambda_1}\Psi_1(x)\frac{1}{2}\left(\theta^{(t)}_1+c\right)+\sqrt{\lambda_2}\Psi_2(x+h)\frac{d}{2}
\end{align*}
Take $a=\frac{1}{2}(\theta^{(t)}_1+c)$ and $b=\frac{d}{2}$. The case for when $\Omega_2$ is active is analogous.
\end{proof}

Therefore, any $\etab^*$ generated by the method is in the space
\[V = \left\{\etab^*; \etab^*(x) = \left\{\begin{array}{ll}
                     \sqrt{\lambda_1}\Psi_1(x)e, & x\in R_1,\\[0.2cm]
                     \frac{1}{2}\left(\sqrt{\lambda_1}\Psi_1(x)a+\sqrt{\lambda_2}\Psi_2(x+h)b\right), & x\in R_2,\\[0.2cm]
                     \frac{1}{2}\left(\sqrt{\lambda_1}\Psi_1(x-h)c+\sqrt{\lambda_2}\Psi_2(x)d\right), & x\in R_3,\\[0.2cm]
                     \sqrt{\lambda_1}\Psi_1(x)g, & x\in R_4,
                     \end{array}\right.\quad a, b, c, d, e, g\in\mathcal{R}\right\}\] 
     
We may then be inclined to say that a global basis can be found for this method, just like in the previous ones. However, \textbf{this algorithm is no longer a Markov chain in} $\thetab$. This is clear in the example above: each new $\etab$ requires the entire history of the accepted values of $\thetab$. As a consequence, the probability of acceptance of the new sample value does not depend on the previous $\thetab$ alone. 

Notice that this is still a Markov chain in $\etab$. We can try to continue the analysis entirely on $\etab$ instead of $\thetab$. But, because the expressions for the $\etab$ in $R_2$ and $R_3$ have previous values of $\thetab$, there are six degrees of freedom in its construction: four as in the theorem, and one for each $R_1$ and $R_4$. In the example, $dim(V)=4$, since there is only one point in $R_2$ and $R_3$, and the two degrees of freedom inside each set will become just one degree of freedom. 
   
\vspace{0.5cm}
\noindent\textit{2. What is the prior distribution on this set V?}

The locally averaged method in \cite{ali2024multiscale} evaluates the prior density in $V$ by \[\pi_{\etab}(\etab^{(t)}):=\pi_{\thetab}(\thetab^{(t)}).\]
While this idea made sense for the previous uncoupled and globally coupled methods, it is not well-defined in this case. The reason why is that the relation between $\thetab^{(t)}$ and $\eta^{(t)}$ is no longer one-to-one. Another way to see it is that $\thetab\in \mathcal{R}^2$, while $\etab$ is in $\mathcal{R}^4$.

\subsubsection{Convergence conditions for the MCMC}    

At a first glance, we can try to use the same instrumental $q_i(\thetab^*|\thetab)$ and its value for an instrumental distribution on $\etab^*|\etab$. However, there is a subtle issue with this approach, which is again caused by the increased degrees of freedom in the space of functions that are simulated. 

Namely, this instrumental distribution cannot sample from all the possible values for $\etab$, and  \textit{Condition 3} --- or its alternative with $\epsilon$ and $\delta$ --- will be violated (see Section \ref{sec:mh}). Whether the entire $\thetab$ is modified at each iteration, or a full Gibbs cycle is performed, those changes span only $\mathcal{R}^2$ and not the four dimensional $V$. We conclude that $q_i(x|y) = 0$ for some $x, y$, even if any distance is chosen in $R^2$ and $|x-y|$ is small.  

\subsubsection{Discussion} We conclude that, regardless if we analyze this method from the point of view of $\etab$ or of $\thetab$, it does not satisfy all the sufficient conditions for convergence. This is still a Markov chain in $\etab$, and it is sampled from the set $V$ that we found. However, convergence is not guaranteed and we don't know yet what its limiting distribution is supposed to be. For this reason, not all of the items of our framework were discussed for this algorithm.

An alternative way to analyze this locally coupled method is to see it as an approximation of another Markov chain. Finding this chain, and determining how close this approximation is, is a future direction for this work.

%%%%%%%%%%%%%%%%%%%%%%%%%%%%%%%%%%%%%%%%%%%%%%%%%%%%%%%%%%%%%%%%%%%%%%%%%%%%%%%%%%%%%%%%
%%%%%%%%%%%%%%%%%%%%%%%%%%%%%%%%%%%%%%%%%%%%%%%%%%%%%%%%%%%%%%%%%%%%%%%%%%%%%%%%%%%%%%%
\section{Conclusion}
\label{sec:conclusion}

This work was the first step towards understanding multiscale sampling methods. We showed how they are a whole family of methods that depend on the way in which initially independent functions can be put together through a operator $G$. 

Next, we proposed a new framework to analyze those methods. Its premise was that different multiscale methods can be understood as simulations from posterior distributions that have the same likelihood function on $\etab$, but different priors. The backbone of our new framework was then in obtaining the properties for the Markov chain in terms of the field $\etab$ and its state space $V$, given the chain in $\thetab$. 

If the one-to-one correspondence between $\etab$ and $\thetab$ is still valid, the analysis is simple and the convergence conditions follow from the convergence of a strictly positive target distribution $f$ and the good properties of a preconditioned Crank-Nicolson step or a random walk sampler. However, while applying this framework to the original multiscale sampler, here called \textit{locally averaged} method, we detected two issues that weren't easy to spot otherwise. 

Two open problems remain. The first is proving the integrability condition that we called \textit{Condition 7}, which is required to guarantee that the Metropolis-within-Gibbs approach constructs a chain that converges for any starting point. The vector of parameters $\thetab$ is mixed in a complicated way by the inverse differential operator, inside the likelihood. We anticipate that proving this condition requires some estimates on elliptic differential operators and also bounds on Gaussian functions, in line with the work of \cite{stuart2010inverse}. 

Secondly, we must find a modification that makes the local averaging idea a valid MCMC method. This is underway; we are working on a simple modification that keeps only one previously accepted value for $\thetab$ and ``forgets'' older ones. However, its analysis is more sophisticated: the previously kept coefficients are ''hidden'' states, and we are working on answering the questions from our framework for this new idea. 

Lastly, the next step in the analysis of Multiscale Sampling Methods is to build upon this framework to perform the mean-squared error analysis between the simulated distribution $f$ on $V$ and the actual distribution on $L^2(\Omega)$.

\section{Acknowledgments} 

This material is based upon work supported by the National Science Foundation under Grant No. 2401945. Any opinions, findings, and conclusions or recommendations expressed in this material are those of the authors and do not necessarily reflect the views of the National Science Foundation. The work of F.P. is also partially supported by The University of Texas at Dallas Office of Research and Innovation through the SPARK program.

\bibliographystyle{elsarticle-num} 
\bibliography{multiscalepaper.bib}

\begin{thebibliography}{10}
\expandafter\ifx\csname url\endcsname\relax
  \def\url#1{\texttt{#1}}\fi
\expandafter\ifx\csname urlprefix\endcsname\relax\def\urlprefix{URL }\fi
\expandafter\ifx\csname href\endcsname\relax
  \def\href#1#2{#2} \def\path#1{#1}\fi

\bibitem{stuart2010inverse}
A.~M. Stuart, Inverse problems: A bayesian perspective, Acta Numerica 19 (2010)
  451--559.
\newblock \href {https://doi.org/10.1017/s0962492910000061}
  {\path{doi:10.1017/s0962492910000061}}.

\bibitem{kaipio2005statistical}
J.~P. Kaipio, E.~Somersalo, Statistical and Computational Inverse Problems,
  Springer New York, 2005.
\newblock \href {https://doi.org/10.1007/b138659} {\path{doi:10.1007/b138659}}.

\bibitem{jaramillo2022towards}
A.~Jaramillo, R.~T. Guiraldello, S.~Paz, R.~F. Ausas, F.~S. Sousa, F.~Pereira,
  G.~C. Buscaglia, Towards hpc simulations of billion-cell reservoirs by
  multiscale mixed methods, Computational Geosciences 26~(3) (2022) 481--501.
\newblock \href {https://doi.org/10.1007/s10596-022-10131-z}
  {\path{doi:10.1007/s10596-022-10131-z}}.

\bibitem{ali2024multiscale}
A.~Ali, A.~Al-Mamun, F.~Pereira, A.~Rahunanthan, Multiscale sampling for the
  inverse modeling of partial differential equations, Journal of Computational
  Physics 497 (2024) 112609.
\newblock \href {https://doi.org/10.1016/j.jcp.2023.112609}
  {\path{doi:10.1016/j.jcp.2023.112609}}.

\bibitem{dodwell2019multilevel}
T.~J. Dodwell, C.~Ketelsen, R.~Scheichl, A.~L. Teckentrup, Multilevel markov
  chain monte carlo, SIAM Review 61~(3) (2019) 509--545.
\newblock \href {https://doi.org/10.1137/19m126966x}
  {\path{doi:10.1137/19m126966x}}.

\bibitem{levin2017markov}
D.~Levin, Y.~Peres, Markov Chains and Mixing Times, American Mathematical
  Society, 2017.
\newblock \href {https://doi.org/10.1090/mbk/107} {\path{doi:10.1090/mbk/107}}.

\bibitem{meyn2009markov}
S.~Meyn, R.~L. Tweedie, P.~W. Glynn, Markov Chains and Stochastic Stability,
  Cambridge University Press, 2009.
\newblock \href {https://doi.org/10.1017/cbo9780511626630}
  {\path{doi:10.1017/cbo9780511626630}}.

\bibitem{casella2002statistical}
G.~Casella, R.~L. Berger, Statistical Inference, second edition Edition,
  Thomson Learning, 2002.

\bibitem{tierney1998note}
L.~Tierney, A note on metropolis-hastings kernels for general state spaces, The
  Annals of Applied Probability 8~(1) (Feb. 1998).
\newblock \href {https://doi.org/10.1214/aoap/1027961031}
  {\path{doi:10.1214/aoap/1027961031}}.

\bibitem{christen2005markov}
J.~A. Christen, C.~Fox, Markov chain monte carlo using an approximation,
  Journal of Computational and Graphical Statistics 14~(4) (2005) 795--810.
\newblock \href {https://doi.org/10.1198/106186005x76983}
  {\path{doi:10.1198/106186005x76983}}.

\bibitem{ali2021multiscale}
A.~Ali, Multiscale sampling for subsurface characterization, Ph.D. thesis, The
  University of Texas at Dallas (2021).

\bibitem{cotter2011variational}
S.~L. Cotter, M.~Dashti, A.~M. Stuart, Variational data assimilation using
  targetted random walks, International Journal for Numerical Methods in Fluids
  68~(4) (2011) 403--421.
\newblock \href {https://doi.org/10.1002/fld.2510}
  {\path{doi:10.1002/fld.2510}}.

\bibitem{cotter2013mcmc}
S.~L. Cotter, G.~O. Roberts, A.~M. Stuart, D.~White, Mcmc methods for
  functions: Modifying old algorithms to make them faster, Statistical Science
  28~(3) (Aug. 2013).
\newblock \href {https://doi.org/10.1214/13-sts421}
  {\path{doi:10.1214/13-sts421}}.

\bibitem{tierney1994markov}
L.~Tierney, Markov chains for exploring posterior distributions, The Annals of
  Statistics 22~(4) (Dec. 1994).
\newblock \href {https://doi.org/10.1214/aos/1176325750}
  {\path{doi:10.1214/aos/1176325750}}.

\bibitem{muller1991generic}
P.~M\"uller, A generic approach to posterior integration and gibbs sampling,
  Tech. rep., Purdue University (1991).

\bibitem{roberts2006harris}
G.~O. Roberts, J.~S. Rosenthal, Harris recurrence of metropolis-within-gibbs
  and trans-dimensional markov chains, The Annals of Applied Probability 16~(4)
  (Nov. 2006).
\newblock \href {https://doi.org/10.1214/105051606000000510}
  {\path{doi:10.1214/105051606000000510}}.

\bibitem{sherlock2010random}
C.~Sherlock, P.~Fearnhead, G.~O. Roberts, The random walk metropolis: Linking
  theory and practice through a case study, Statistical Science 25~(2) (May
  2010).
\newblock \href {https://doi.org/10.1214/10-sts327}
  {\path{doi:10.1214/10-sts327}}.

\bibitem{alexanderian2017brief}
A.~Alexanderian, A brief note on the karhunen-loève expansion (2015).
\newblock \href {http://arxiv.org/abs/1509.07526v2}
  {\path{arXiv:1509.07526v2}}.

\bibitem{ghanem2003stochastic}
R.~G. Ghanem, P.~D. Spanos, Stochastic finite elements: a spectral approach,
  revised edition Edition, Dover, 2003.

\bibitem{ossiander2014conditional}
M.~E. Ossiander, M.~Peszynska, V.~S. Vasylkivska, Conditional stochastic
  simulations of flow and transport with karhunen-loève expansions, stochastic
  collocation, and sequential gaussian simulation, Journal of Applied
  Mathematics 2014 (2014) 1--21.
\newblock \href {https://doi.org/10.1155/2014/652594}
  {\path{doi:10.1155/2014/652594}}.

\bibitem{schabenberger2017statistical}
O.~Schabenberger, C.~A. Gotway, Statistical Methods for Spatial Data Analysis:
  Texts in Statistical Science, Chapman and Hall/CRC, 2017.
\newblock \href {https://doi.org/10.1201/9781315275086}
  {\path{doi:10.1201/9781315275086}}.

\bibitem{christakos1992random}
G.~Christakos, Random Field Models in Earth Sciences, Elsevier, 1992.
\newblock \href {https://doi.org/10.1016/c2009-0-22238-0}
  {\path{doi:10.1016/c2009-0-22238-0}}.

\bibitem{bear1987modeling}
J.~Bear, A.~Verruijt, Modeling Groundwater Flow and Pollution, Springer
  Netherlands, 1987.
\newblock \href {https://doi.org/10.1007/978-94-009-3379-8}
  {\path{doi:10.1007/978-94-009-3379-8}}.

\bibitem{guiraldello2018multiscale}
R.~T. Guiraldello, R.~F. Ausas, F.~S. Sousa, F.~Pereira, G.~C. Buscaglia, The
  multiscale robin coupled method for flows in porous media, Journal of
  Computational Physics 355 (2018) 1--21.
\newblock \href {https://doi.org/10.1016/j.jcp.2017.11.002}
  {\path{doi:10.1016/j.jcp.2017.11.002}}.

\bibitem{xu2024domain}
Z.~Xu, Q.~Liao, J.~Li, Domain-decomposed bayesian inversion based on local
  karhunen-loève expansions, Journal of Computational Physics 504 (2024)
  112856.
\newblock \href {https://doi.org/10.1016/j.jcp.2024.112856}
  {\path{doi:10.1016/j.jcp.2024.112856}}.

\end{thebibliography}

\end{document}